\documentclass[twocolumn,aps,physrev,
superscriptaddress,
floatfix,longbibliography,amsmath,amssymb]{revtex4}
\usepackage{graphicx}% Include figure files
\usepackage{dcolumn}% Align table columns on decimal point
\usepackage{setspace}
\usepackage{bm}% bold math
\usepackage{multirow}
\usepackage[colorlinks,linkcolor=blue,anchorcolor=blue,citecolor=blue,urlcolor=blue]{hyperref}
\usepackage[utf8]{inputenc}
\usepackage{threeparttable}
\usepackage{multirow,booktabs}
\usepackage{orcidlink}
\usepackage{comment}

\newcommand{\MCPUS}{MCP_{US}}
\newcommand{\MCPTGT}{MCP_{TGT}}
\newcommand{\MCPDS}{MCP_{DS}}
\begin{document}

\preprint{APS/123-QED}

\title{Deconstructing the emission order of protons, neutrons and $\alpha$-particles following fusion in $^{28,30,32}$Si + $^{28}$Si}

\author{Rohit Kumar\,\orcidlink{0000-0002-0450-7218}}
\affiliation{%
Department of Chemistry and Center for Exploration of Energy and Matter, Indiana University\\
2401 Milo B. Sampson Lane, Bloomington, Indiana 47408, USA}%

\author{H. Desilets\,\orcidlink{0009-0009-4348-6526}}
\affiliation{%
Department of Chemistry and Center for Exploration of Energy and Matter, Indiana University\\
2401 Milo B. Sampson Lane, Bloomington, Indiana 47408, USA}%

\author{J.~E. Johnstone}
\affiliation{%
Department of Chemistry and Center for Exploration of Energy and Matter, Indiana University\\
2401 Milo B. Sampson Lane, Bloomington, Indiana 47408, USA}%

\author{S. Hudan\,\orcidlink{0000-0002-9722-2245}}
\affiliation{%
Department of Chemistry and Center for Exploration of Energy and Matter, Indiana University\\
2401 Milo B. Sampson Lane, Bloomington, Indiana 47408, USA}%

\author{D. Chattopadhyay}
\affiliation{%
Department of Chemistry and Center for Exploration of Energy and Matter, Indiana University\\
2401 Milo B. Sampson Lane, Bloomington, Indiana 47408, USA}%
\affiliation{%
Department of Physics, The ICFAI University Tripura, West Tripura 799210, India}%

\author{R.~T. deSouza\,\orcidlink{0000-0001-5835-677X}}%
\email[Corresponding author: ] {desouza@iu.edu}
\affiliation{%
Department of Chemistry and Center for Exploration of Energy and Matter, Indiana University\\
2401 Milo B. Sampson Lane, Bloomington, Indiana 47408, USA}%

\author{D. Ackermann}
\affiliation{%
GANIL, CEA/DRF-CNRS/IN2P3, \\
Blvd. Henri Becquerel, F-14076, Caen, France}%

\author{M. Basson}
\affiliation{%
Facility for Rare Isotope Beams, Michigan State University, East Lansing, MI 48823, USA}%

\author{K.~W. Brown}
\affiliation{%
Department of Chemistry, Michigan State University, East Lansing MI 48824 USA}%

\author{A. Chbihi\,\orcidlink{0000-0001-5653-4325}}%
\affiliation{%
GANIL, CEA/DRF-CNRS/IN2P3, \\
Blvd. Henri Becquerel, F-14076, Caen, France}%

\author{K.~J. Cook\,\orcidlink{0000-0002-5911-1333}}%
\affiliation{%
Facility for Rare Isotope Beams, Michigan State University, East Lansing, MI 48823, USA}%
\affiliation{%
Department of Nuclear Physics and Accelerator Applications, Research School of Physics, Australian National University, Canberra, ACT 2601, Australia}%

\author{M. Famiano}
\affiliation{%
Department of Physics, Western Michigan University, Kalamazoo, Michigan 49008, USA}%

\author{T. Genard}
\affiliation{%
GANIL, CEA/DRF-CNRS/IN2P3, \\
Blvd. Henri Becquerel, F-14076, Caen, France}%

\author{I.~M. Harca}
\affiliation{%
Facility for Rare Isotope Beams, Michigan State University, East Lansing, MI 48823, USA}%

\author{S.~N. Paneru}
\affiliation{%
Facility for Rare Isotope Beams, Michigan State University, East Lansing, MI 48823, USA}%
\affiliation{%
Los Alamos National Laboratory, Los Alamos, NM 87545, USA}%

%end of authors
%%%%%%%%%%%%%%%%%%%%%%%%%%%%%%%%%%%%%%%%%%%

\date{\today}
\begin{abstract}
A high-quality measurement of proton and $\alpha$-particle emission associated with fusion of $^{28,30,32}$Si with a $^{28}$Si target is described. Evaporation residues produced by de-excitation of the compound nucleus were identified by an energy time-of-flight (ETOF) measurement while emitted light-charged particles were identified using the $\Delta$E-E technique. Comparison of the experimentally measured  charged particle multiplicities and energy spectra with the predictions of the statistical decay model code, GEMINI++, allows one to deduce interesting details of the de-excitation cascade and its dependence on neutron-excess. The impact of modifying the sequence of particle emissions on the average energy and multiplicity is examined.

\end{abstract}
\maketitle

%\tableofcontents

\section{Introduction}

Clusters play an important role in both the structure and reactions of nuclei. In the case of heavy nuclei, the presence of clusters in the nuclear medium is evident through cluster radioactivity \cite{Rose84} while for light nuclei it can be an intrinsic part of their structure \cite{Horiuchi10}.  The archetypal case of the importance of clusters in nuclear reactions is the Hoyle state responsible for $^{12}$C production in red giant stars \cite{Hoyle54, Cook57}. In Fermionic Molecular Dynamics calculations the Hoyle state is found to be dominated by dilute $\alpha$-cluster configurations \cite{Neff08}. While it is well established that cluster structure is intrinsic to the nature of nuclei, several key questions are unanswered. Although the existence of an $\alpha$-condensate is favored for nuclear matter at low density \cite{Ropke13, Typel10}, the dependence of this condensate on the excitation/temperature of the medium is yet to be established. Is formation of $\alpha$-clusters strongly suppressed as the medium becomes increasingly neutron-rich? The latter question is of particular relevance to the structure and reactions of nuclei in neutron-rich environments such as the outer-crusts of neutron stars or the ejecta in neutron star mergers. Although $\alpha$-clustering in self-conjugate (N=Z) nuclei has received the most attention, clustering in non-self-conjugate nuclei has been the focus of some recent studies \cite{Avila14}. 

It was recently observed that the yield for $\alpha$-emission following fusion of $^{18}$O with $^{12}$C nuclei significantly exceeded the predictions of the statistical model \cite{Vadas15}. 
Although the yield was under-predicted, the measured energy and angular distributions of the emitted $\alpha$-particles are well described by a fusion-evaporation process. 
The 4-5 fold excess of $\alpha$-particles as compared to  standard statistical model calculations \cite{PACE4} was interpreted as the persistence of an $\alpha$-cluster structure in the entrance channel through the fusion process. A similar enhancement of $\alpha$-particles relative to the statistical model expectation was also observed for fusion of $^{16}$O + $^{12}$C \cite{Tabor77} and $^{16}$O + $^{13}$C \cite{Papa86}.

Calculation of clustering in the fusion of light and mid-mass nuclei indicates that $\alpha$-clusters could also be dynamically generated during the fusion process \cite{Schuetrumpf17}. These calculations indicate not only the formation of cluster states during the collision but their evolution on the timescale of the rotational period. The observed $\alpha$-cluster states arise from density oscillations introduced by the collision dynamics which produce nuclear matter at low density. 
It should be emphasized that the TD-DFT/NLF calculations do not predict the emission of $\alpha$-particles, only the existence of these cluster states in the pre-compound system. Nevertheless, the calculations do demonstrate that large amplitude collective motion of the pre-compound system is not simply described by the vibration of two nucleonic fluids but is far more complex. 

A proper understanding of cluster emission, whether dynamical or statistical in origin, requires an accurate measurement of the competing nucleon emission. 
In this work we describe a high-quality measurement of proton and $\alpha$-particle emission associated with fusion of $^{28,30,32}$Si with a $^{28}$Si target. Following fusion, the compound nucleus de-excites predominantly by emission of protons, neutrons, and $\alpha$-particles. Comparison of the experimentally measured multiplicities and energy spectra of the charged particles with a statistical decay model allows one to deduce interesting details of the de-excitation cascade. Comparison of the different systems enables examination of how the de-excitation is impacted by the neutron-excess of the compound nucleus. This improved understanding of standard statistical emission of nucleons and $\alpha$-particles is necessary in order to probe the dynamical emission of the $\alpha$-clusters.

\section{Experimental details}
The experiment was conducted at Michigan State University's  National Superconducting Cyclotron Laboratory using the ReA3 linear accelerator \cite{Villari23}. Although ReA3 was primarily developed to serve as a re-accelerator for radioactive beams produced at the Facility for Rare Isotope Beams (FRIB), it is also capable of operating in stand-alone mode. In this mode, it can accelerate either stable or long-lived radioactive ions delivered from an ion source. This capability allowed it to deliver the beams of both stable $^{28,30}$Si and radioactive $^{32}$Si (t$_{1/2}$ = 163.3 y) nuclei. After their generation in an ion source, ions are charge-bred in the electron beam ion trap (EBIT), then extracted with charge selection and accelerated to $\sim$3 MeV/A with an acceleration frequency of 80.5 MHz. The accelerated beams impinged on a $^{28}$Si target with a beam intensity of $\sim$10$^{5}$ ions/s.

\begin{figure}[h]
\begin{center}
\includegraphics[scale=0.56]{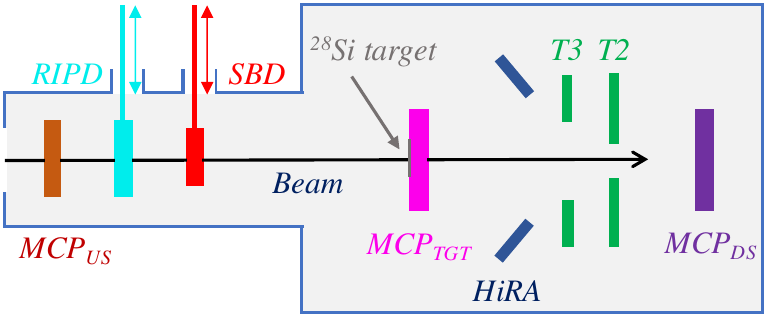}
\caption{Schematic of the experimental setup used to measure fusion in $^{28,30,32}$Si + $^{28}$Si. Shown are the elements used for the beam identification: the three microchannel plate detectors 
$\MCPUS, \MCPTGT $, and $\MCPDS $
together with the SBD and RIPD detector. Identification of reaction products is accomplished using the $\MCPTGT$ with the annular Si detectors (T2,T3), and the HiRA telescopes.}

\label{fig:Setup}
\end{center}
\end{figure}

\begin{figure}[h]
\begin{center}
\includegraphics[scale=0.35]{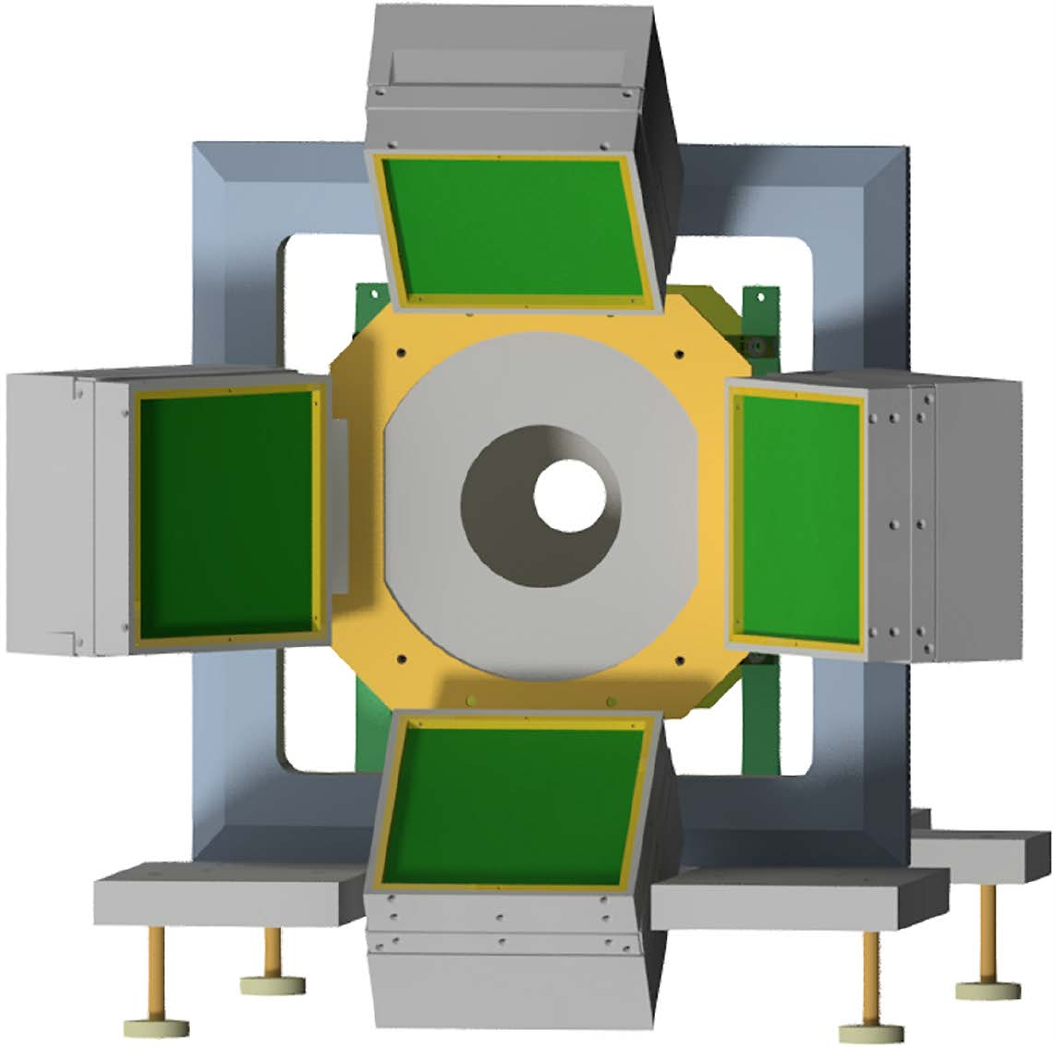}
\caption{CAD of the compact arrangement of the four HiRA telescopes together with the annular Si detectors T2 and T3 used to detect the ERs.
}
\label{fig:CAD}
\end{center}
\end{figure}

A schematic of the experimental setup which provided efficient detection of the fusion products is presented in Fig.~\ref{fig:Setup}. The beam identification portion of the setup consisted of three microchannel plate (MCP) detectors \cite{Bowman78,Steinbach14} and an axial-field ionization chamber designated Rare-Ion Purity Detector (RIPD) \cite{Vadas16}. Periodic insertion of a surface barrier Si detector (SBD) into the beam path immediately downstream of RIPD provided a measure of the beam energy incident on the target. Heavy fusion products, produced by de-excitation of the compound nucleus, were detected in the angular range 3.6$^\circ\leq\theta_{lab}\leq$ 10.7$^\circ$ and  $11.0^\circ\leq\theta_{lab}\leq 21.2^\circ$ using two segmented annular silicon detectors designated T2 and T3 in Fig.~\ref{fig:Setup}. Light-charged particles (Z$\leq$2), principally protons and $\alpha$-particles, were identified in the angular range $32.6^o\leq\theta_{lab}\leq 68.4^o$ by four HiRA telescopes \cite{Wallace07}. The compact arrangement of these telescopes and the T2 and T3 detectors is depicted in Fig.~\ref{fig:CAD}.

The first element of the experimental setup in the beam path is the $\MCPUS$ detector. 
This detector utilizes a ExB design \cite{Bowman78,Steinbach14} with a thin 20 $\mu$g/cm$^2$ carbon foil to provide a timing signal for the passage of an incident ion with minimal energy loss. This detector has an intrinsic time resolution of $\sim$200ps. 
Approximately 45 cm downstream of the $\MCPUS$ detector, the beam passes through a compact axial-field ionization chamber designated RIPD that was used to identify beam contaminants. 
As an ion in the beam passes through RIPD, it deposits an energy, $\Delta$E, characteristic of its identity consequently altering its velocity. 
Fast charge collection in RIPD allows it to operate at a high beam rate of 10$^5$ ions/s \cite{Vadas16}.

Approximately 180 cm downstream of the $\MCPUS$, the 
beam impinged on a second 
ExB design MCP detector, designated $\MCPTGT$. 
The secondary emission foil in this detector, 215$\pm$10 $\mu$g/cm$^2$-thick $^{28}$Si, also served as the target. These two MCPs provided a time-of-flight (TOF) measurement as well as a count of the total beam particles incident on the target. The third ExB MCP detector located downstream of the target $\MCPDS$, aided with the rejection of un-reacted beam.

Signals from all detectors were processed through standard analog electronics. For the silicon detectors and RIPD signals this consisted of high-quality charge-sensitive amplifiers \cite{zepto}, as well as shaping and timing-filter amplifiers. The energy signals were then digitized in peak-sensing ADCs (Caen V785). Timing signals were amplified, discriminated and then digitized in time-to-digital converters (Caen V1290). Readout of the digitized signals by a VM-USB based data acquisition was triggered by either detection of a particle in  one of the annular silicon detectors or in one of the four HiRA telescopes.

\begin{figure}[h]
\begin{center}
\includegraphics[scale=0.37]{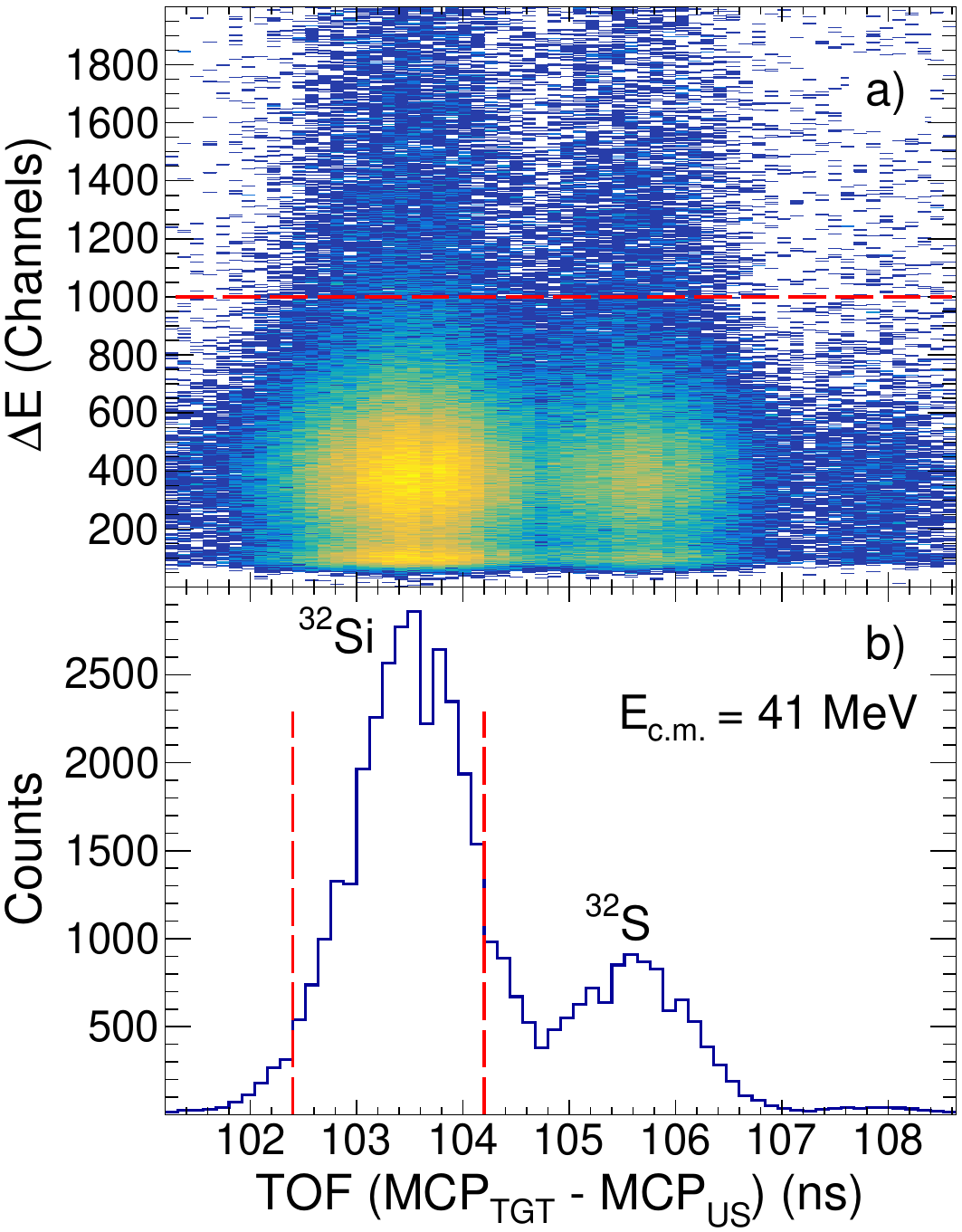}
\caption{A PID spectrum using RIPD and two MCP detectors. a) $\Delta$E vs TOF spectrum for $^{32}$Si beam on target at $E_{c.m.}$ = 41 MeV. The 2nd peak at higher TOF is $^{32}$S which was predicted as a contaminant. A energy threshold at $\sim$1000 channel provides a means to reject pile-up. b) Time projection of the data in panel a). Selection of an incident $^{32}$Si ion is indicated by the vertical dashed lines. 
}
\label{fig:BeamPID}
\end{center}
\end{figure}

Together with the $\MCPUS$ and $\MCPTGT$ detectors, RIPD provided identification of incident particles based on their $\Delta$E-TOF. A representative $\Delta$E-TOF spectrum is presented in Fig.~\ref{fig:BeamPID}a for $^{32}$Si (t$_{1/2}$=163.3 y) beam at $E_{c.m.}$ = 41 MeV incident on the $^{28}$Si target. Two peaks well separated in TOF can be seen where the peak at higher TOF corresponds to a $^{32}$S contaminant present with the $^{32}$Si beam. While the energy signal allows us to reject pile-up, as indicated by the horizontal dashed line,  the principal rejection of beam contaminants is made on the basis of time-of-flight. The selection of $^{32}$Si incident ions is depicted in Fig.~\ref{fig:BeamPID}b by the vertical dashed lines.

Following fusion of an incident ion with a $^{28}$Si target nucleus, the resulting compound nucleus de-excites via particle emission. Due to momentum conservation this particle emission  deflects the resulting evaporation residue (ER) away from the beam direction. ERs were detected by two annular silicon detectors designated T2 and T3 in Fig.~\ref{fig:Setup}. The T2 and T3 detectors are S9 and S1 designs respectively from Micron Semiconductor \cite{MicronSemiconductor}. Each of these detectors is a reverse-biased pn junction segmented into 16 pie-shaped sectors on the ohmic side and concentric rings on the junction side. This segmentation provides angular information for the detected particles. In addition to providing an energy signal the annular detectors also provided a fast timing signal. This signal allowed sub-nanosecond timing to be performed which was critical for distinguishing the ERs from un-reacted beam \cite{deSouza11}. The geometric efficiency of these two detectors was determined by examining the the fraction of ERs predicted by the statistical decay code GEMINI++ to lie within the angular acceptance. Combined these two detectors provided a geometric efficiency of $\sim$56-62$\%$ for detection of the ER. Multiple beams during the experiment provided elastic scattering peaks that were used to energy calibrate the T2 and T3 detectors. 

Emitted light-charged particles (LCPs: Z$\leq$2) were detected using four HiRA, $\Delta$E-E-CsI(Tl) telescopes \cite{Wallace07}, as depicted in Fig.~\ref{fig:CAD}. The first element of the telescope is a 65 $\mu m$-thick $\Delta$E single-sided detector with 32 strips. Situated behind it is a 1.5 mm-thick double-sided E detector with 32x32 strips. The segmentation of these detectors allows determination of the detected particle's position. Following the E detector are four 1 cm-thick CsI(Tl) crystals with photodiode (PD) readout. These CsI(Tl)/PD detectors serve to detect high-energy protons that punch-through the E detector. As only the highest energy protons, comprising $<$2$\%$ of the proton yield, were detected in the CsI(Tl)/PD detectors, the analysis utilized only particles stopping in the 1.5 mm-thick E Si detector. The low particle multiplicity in the reactions studied enabled reduction of the electronic channels required by joining adjacent strips in both the $\Delta$E and E detectors. In the $\Delta$E detector, dictated by the detector capacitance, four adjacent strips were coupled to yield 8 strips. In the case of the thick E detector, sixteen adjacent strips were coupled to provide effectively 2 strips on the junction side and 2 orthogonal strips on the ohmic side. This coupling of strips in HiRA resulted in measurement of emitted particles at only a few laboratory angles. As with the determination of the ER detection efficiency, GEMINI++ was used to determine both the geometric efficiency for proton and $\alpha$-particle detection as well as their coincidence efficiency with an ER. The efficiency for detecting protons and $\alpha$-particles is $\sim$14.2-14.6$\%$ and $\sim$16.2-17.2$\%$ respectively. Coincident detection of a proton and an ER has an efficiency of $\sim$7.6-8.9$\%$ while coincident detection of an $\alpha$-particle and ER has an efficiency of $\sim$15.0-16.0$\%$. Use of GEMINI++ to determine the geometric efficiency of the experimental setup both for emitted particles as well as ERs assumes that the angular and energy distributions of emitted particles in GEMINI++ accurately describes the experimental data. The reasonableness of this assumption is subsequently shown.

To perform an energy calibration of the HiRA telescopes the $\alpha$-emitting radioactive sources $^{148}$Gd and $^{232}$U were utilized. Both the $\Delta$E and E detectors in HiRA were calibrated in the same manner. The use of $^{148}$Gd provides a single-line $\alpha$-particle at 3.182 MeV while the $^{232}$U source provides multiple $\alpha$ energies between 5.685 and 8.784 MeV. The 8.784 MeV $\alpha$-particle is particularly useful as it lies close to the punch-though energy of the $\alpha$-particle through the 65 $\mu$m  $\Delta$E detector ($\sim$9 MeV). 

\section{Evaporation Residues}

\begin{figure}
\begin{center}
\includegraphics[scale=0.41]{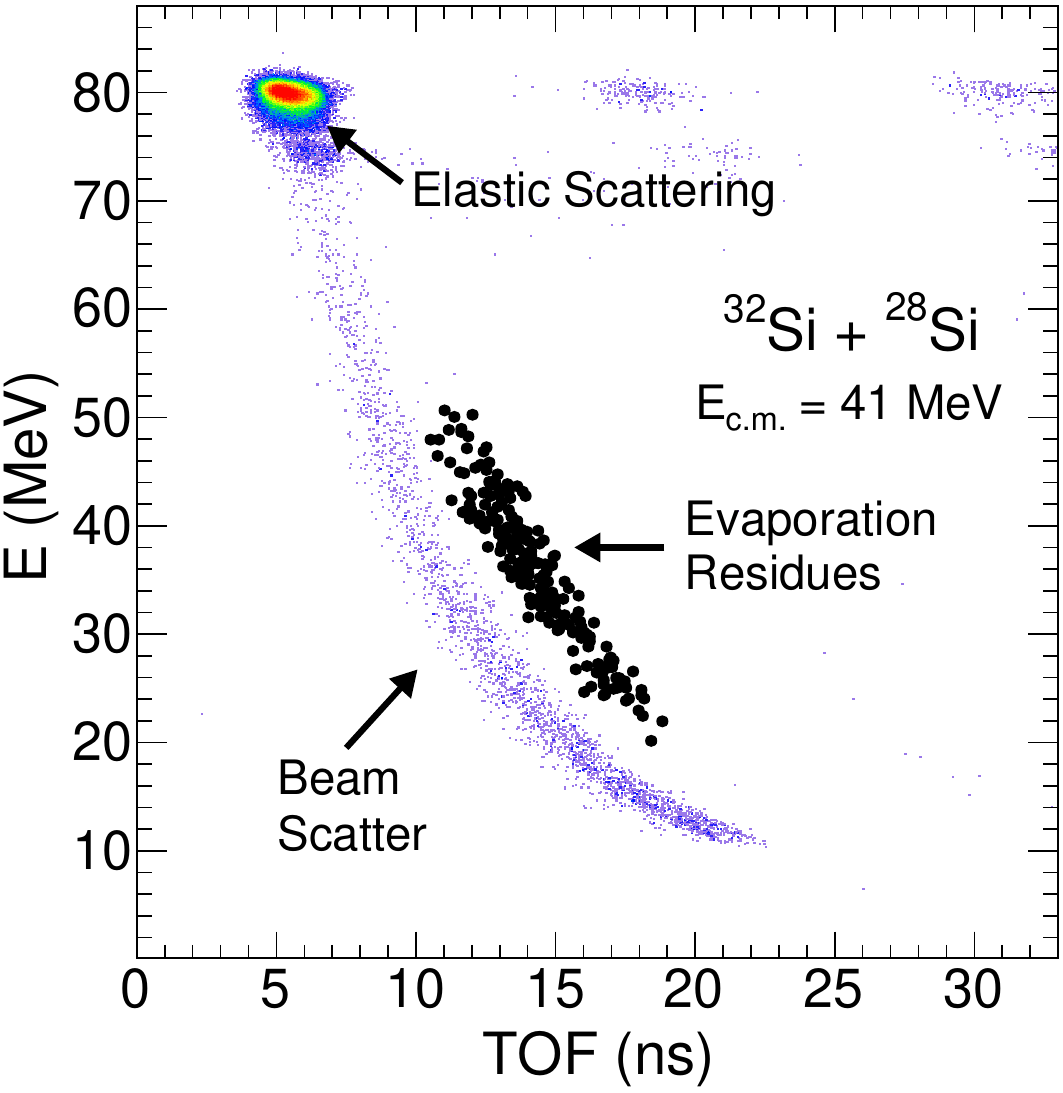}
\caption{Energy vs Time-of-flight (TOF) spectrum for $^{32}$Si beam incident on $^{28}$Si target at $E_{c.m.}$ = 41 MeV.
}
\label{fig:T2ETOF}
\end{center}
\end{figure}

Presented in Fig.~\ref{fig:T2ETOF} is an ETOF spectrum for $^{32}$Si + $^{28}$Si at $E_{c.m.} =$ 41 MeV. The energy (E) corresponds to the energy deposited in the T2 silicon detector and the TOF is the time difference between T2 and $\MCPTGT$. The most prominent feature of this spectrum is an intense peak at $\sim$80 MeV and $\sim$5 ns which corresponds to the elastically scattered beam particles. Visible to the right of the elastic peak and spaced approximately 12 ns apart are the elastic peaks associated with adjacent beam bursts. The tail from the elastic peak extending down in energy with increasing TOF corresponds to slit-scattered beam particles. The slit-scatter line corresponds to ions with the same mass as the beam. Clearly separated from the slit-scatter line and situated to slightly larger TOF an island is observed. For a given energy, the larger TOF observed for this island is quantitatively consistent with the larger mass associated with ERs following fusion. 

\begin{figure}
\begin{center}
\includegraphics[scale=0.42]{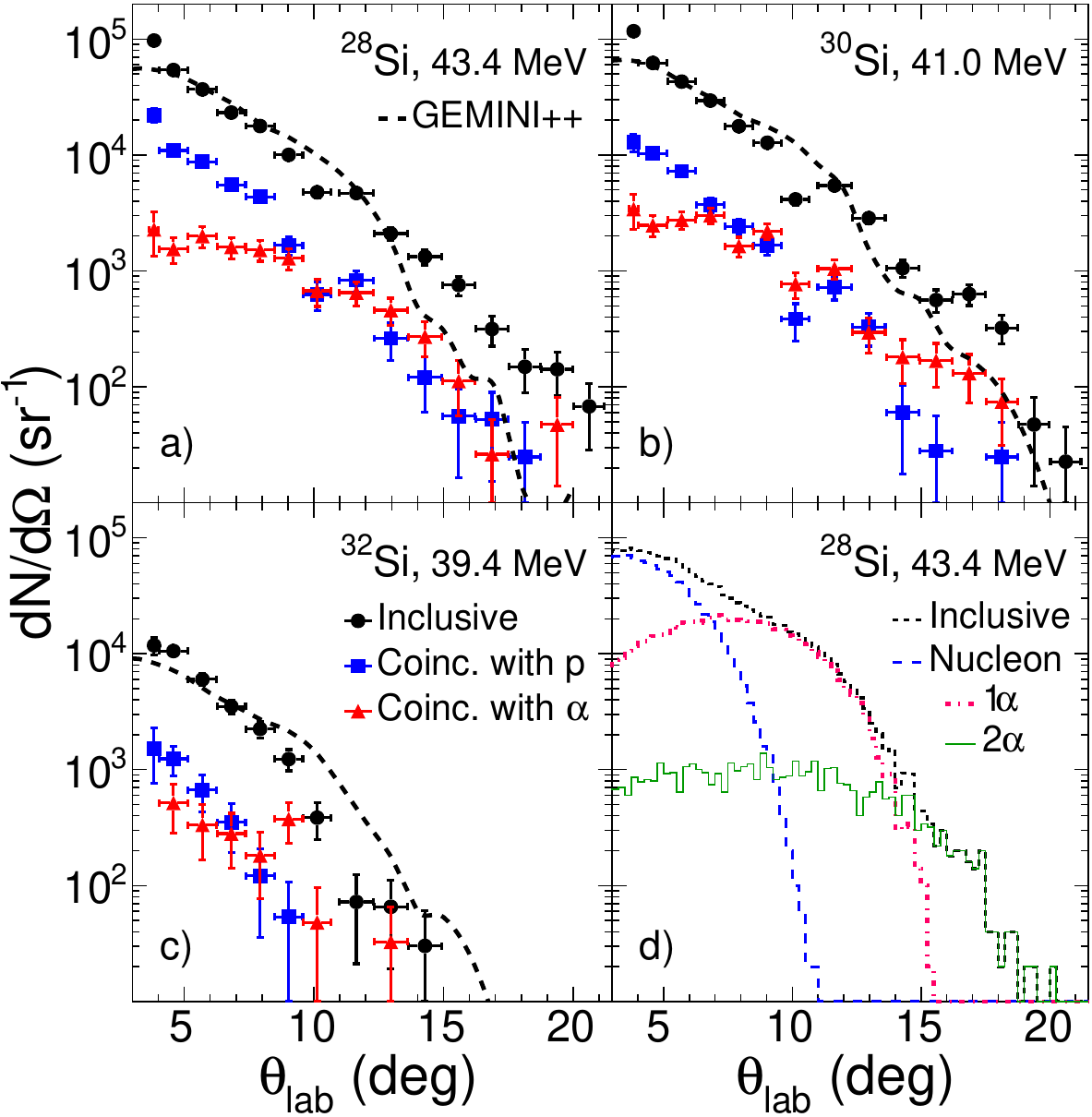}
\caption{Panels a-c) Angular distribution of evaporation residues (ERs) in the laboratory frame for $^{28,30,32}$Si + $^{28}$Si. Both the inclusive ER distribution as well as the distributions coincident with detection of a proton or an $\alpha$-particle are shown. Symbols represent the experimental data while the dashed line indicates the angular distributions predicted by the GEMINI++ model. Error bars shown depict the statistical errors. d) Angular distributions predicted by the GEMINI++ model for different de-excitation channels: Inclusive (dotted black histogram) nucleons only (dashed blue), one $\alpha$ (dashed-dotted red) and two $\alpha$s (solid green). 
}
\label{fig:ERTheta}
\end{center}
\end{figure}

The angular distribution of detected ERs is presented in Fig.~\ref{fig:ERTheta}a,b,c for $^{28}$Si, $^{30}$Si and $^{32}$Si respectively. 
For each system, although only one energy is presented for simplicity the angular distributions for the other incident energies exhibit similar behavior. The inclusive ER distributions are represented by the filled (black) circles. These  distributions are strongly peaked at small angles and extend out to $\theta_{lab}$$\sim$20$^\circ$ indicating that on average only a small transverse momentum has been imparted to the ER. The angular distribution of ERs coincident with detection of a proton, indicated by the filled (blue) squares, is similar to the inclusive one for all three systems. This result suggests that the transverse momentum imparted by an emitted proton is typical of all fusion reactions. 
The ER angular distribution associated with coincident detection of an $\alpha$-particle, depicted by red triangles, is somewhat different. In the case of $^{28}$Si and $^{30}$Si a suppression of the yield at small angles $\theta_{lab}<10^\circ$ is observed.  For the case of $^{32}$Si, limited statistics preclude a definitive conclusion. This suppression of small angle yield can be understood as due to the larger transverse recoil imparted to the ER by the emission of a single $\alpha$-particle as compared to a proton. Emission of additional particles mitigates the effect of the recoil.

To understand the statistical decay component of the proton and $\alpha$-particle yield, we have examined the angular distribution of ERs predicted by the statistical decay model GEMINI++ \cite{Charity10}. GEMINI++ describes the binary, sequential decay of an excited compound nucleus within the Hauser-Feshbach formalism. Decay widths for charged particle and neutron emission are calculated. For the low-excitation and low (Z,A) of the compound nucleus considered the decay is dominated by neutron, proton, and $\alpha$-particle emission. For the relatively low-excitation energy of the compound nucleus (E$^*$$\sim$57 MeV), the mean time between emissions is long validating use of the Bohr independence hypothesis in the statistical model code. The level density parameter for the default calculations is taken as A/7. A maximum angular momentum, $\ell_{max}$, was taken based upon the Bass fusion model \cite{Bass74} with a diffuseness of 2$\hbar$. GEMINI++ accounts for the impact of a barrier distribution on the energy spectra of emitted particles. The barrier distribution can arise from large thermal fluctuations, shape fluctuations, or fluctuations in the diffuseness of the nuclear surface \cite{Charity10}.

The angular distribution predicted by GEMINI++ for all ERs is indicated by the dashed-line in Fig.~\ref{fig:ERTheta}a,b,c.   While the overall description of the shape of the angular distribution by GEMINI++ is generally good, some points are noteworthy. For $^{28}$Si
the model under-predicts the experimental angular distribution at larger angles. In contrast, for $^{32}$Si GEMINI++ overpredicts the yield at large angles with the case of $^{30}$Si being intermediate. As the observation of ERs at large angles is preferentially  associated with $\alpha$-emission (red triangles), this trend can be interpreted as an underestimation of $\alpha$ emission by GEMINI++ for $^{28}$Si and an overprediction of $\alpha$-emission for neutron-rich systems.

A better understanding of the inclusive ERs distribution can be achieved by examination of the predictions of the statistical model code GEMINI++ presented in Fig.~\ref{fig:ERTheta}d. De-excitation of the compound nucleus proceeds via emission of protons, neutrons, and $\alpha$-particles. Emission of only nucleons imparts less transverse momentum to the ER, contributing primarily to the yield of ERs at small angles as indicated by the blue dashed line. In contrast, emission of an $\alpha$-particle populates on average larger angles while suppressing small angles as evident from the red dash-dot line. It is interesting to note that the observation of ER at the largest angles $\theta_{lab}$$>$16$^\circ$ is associated with the emission of two $\alpha$-particles (green solid line) indicating that the balance of different emission channels is encoded in the detailed shape of the angular distribution. 

\begin{figure}
\begin{center}
\includegraphics[scale=0.45]{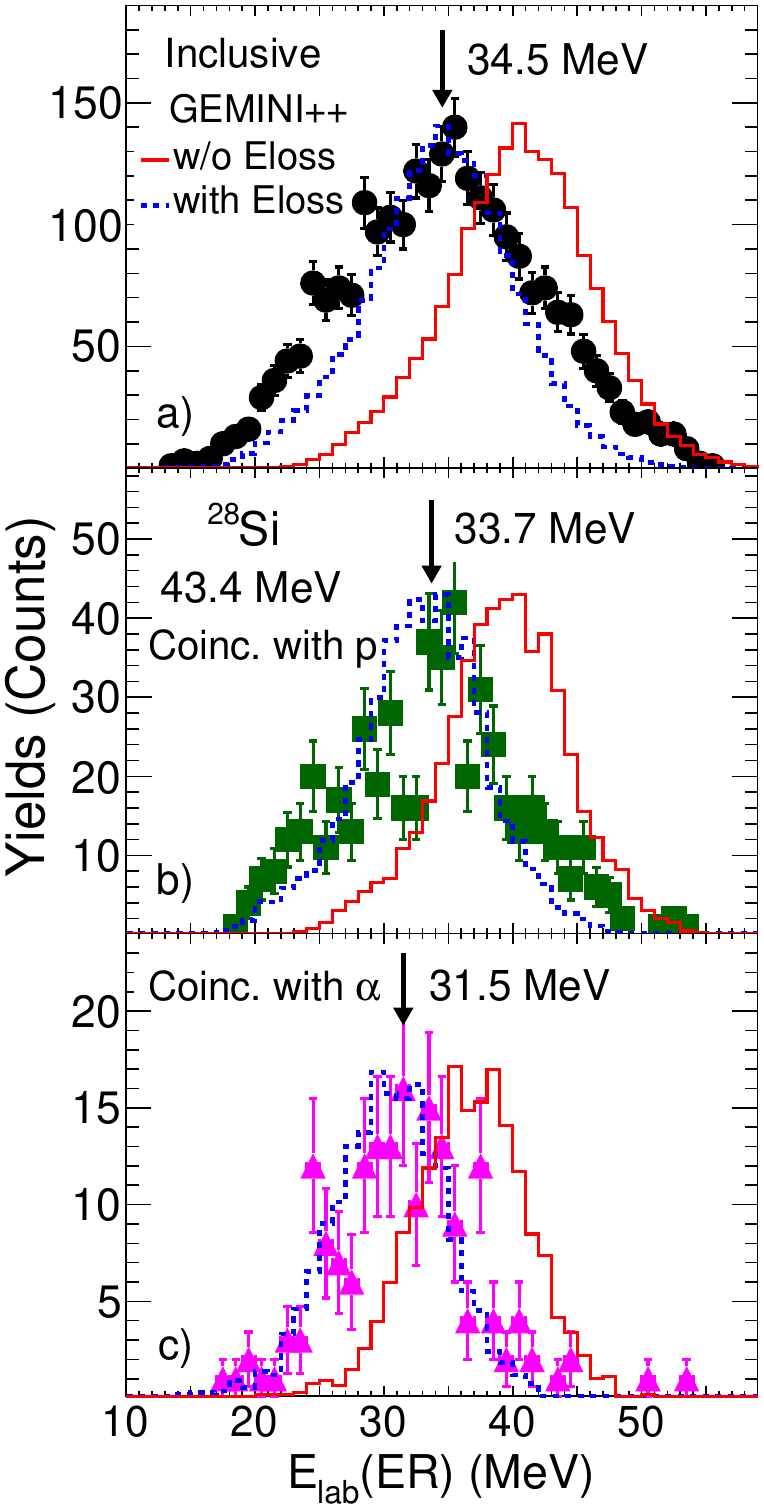}
\caption{Energy distribution of ERs in laboratory frame from $^{28}$Si + $^{28}$Si both inclusive (panel a) as well as with coincident protons and alphas (panels b and c respectively). Statistical errors are represented by the error bars shown. Also shown are the GEMINI++ predictions of the ER energy distributions before and after dead layer corrections.}
\label{fig:EREnergy}
\end{center}
\end{figure}

Depicted in Fig.~\ref{fig:EREnergy} are the laboratory energy distribution of ERs in $^{28}$Si + $^{28}$Si at $E_{c.m.}$ = 43.4 MeV. The mean value of each distribution is represented by a vertical arrow. In Fig.~\ref{fig:EREnergy}a the inclusive ER experimental data is single peaked with a mean value of $E_{lab}$ = 34.5 MeV. The energy distribution predicted by GEMINI++, filtered for angular acceptance of T2 and T3 detectors, is indicated by the solid line (red) and is situated at slightly higher energy ($\langle E \rangle$= 40.5 MeV). Most of the difference between the GEMINI++ predictions and the experimental data can be understood as due to the energy loss \cite{SRIM} in the target and dead layer at the front surface of the annular silicon detectors. Accounting for the energy loss in half the target thickness and a dead layer (0.7 $\mu$m Si equivalent) results in the dashed (blue) line ($\langle E \rangle$= 34.4 MeV) which is in good agreement with the experimental distribution although it is slightly narrower. This difference in width is likely due to the fact that only the (Z,A) of the most probable ER was used in the energy loss calculations.
%{\bf RMS for GEMINI++: 5.724 MeV}

In Fig.~\ref{fig:EREnergy}b,c the ER energy distributions associated with coincident detection of a proton or an $\alpha$-particle are presented. The average energy for events with a coincident proton or $\alpha$-particle is 33.7 MeV and 31.5 MeV respectively, slightly lower than the 34.5 MeV observed for the inclusive distribution. This shift in the average energy occurs because in the experimental setup the proton or $\alpha$-particle is detected in the forward direction imparting a backward recoil to the ER thus lowering its energy. 
Similar to Fig.~\ref{fig:EREnergy}a, the GEMINI++ model predictions are depicted by solid (red) and dashed (blue) lines. As anticipated from the inclusive energy distributions, the solid (red) distributions are situated at slightly higher energies than the experimental data with mean values of $\langle E \rangle$= 39.3 MeV for protons and $\langle E \rangle$= 36.7 MeV for $\alpha$-particles. After incorporating the effect of energy loss, the dashed (blue) distributions with mean values of $\langle E \rangle$= 32.9 MeV for protons and $\langle E \rangle$= 30.4 MeV for $\alpha$-particles, are in reasonable agreement with the experimental data.

\section{Light-charged particle emission}

\begin{figure}
\begin{center}
\includegraphics[scale=0.42]{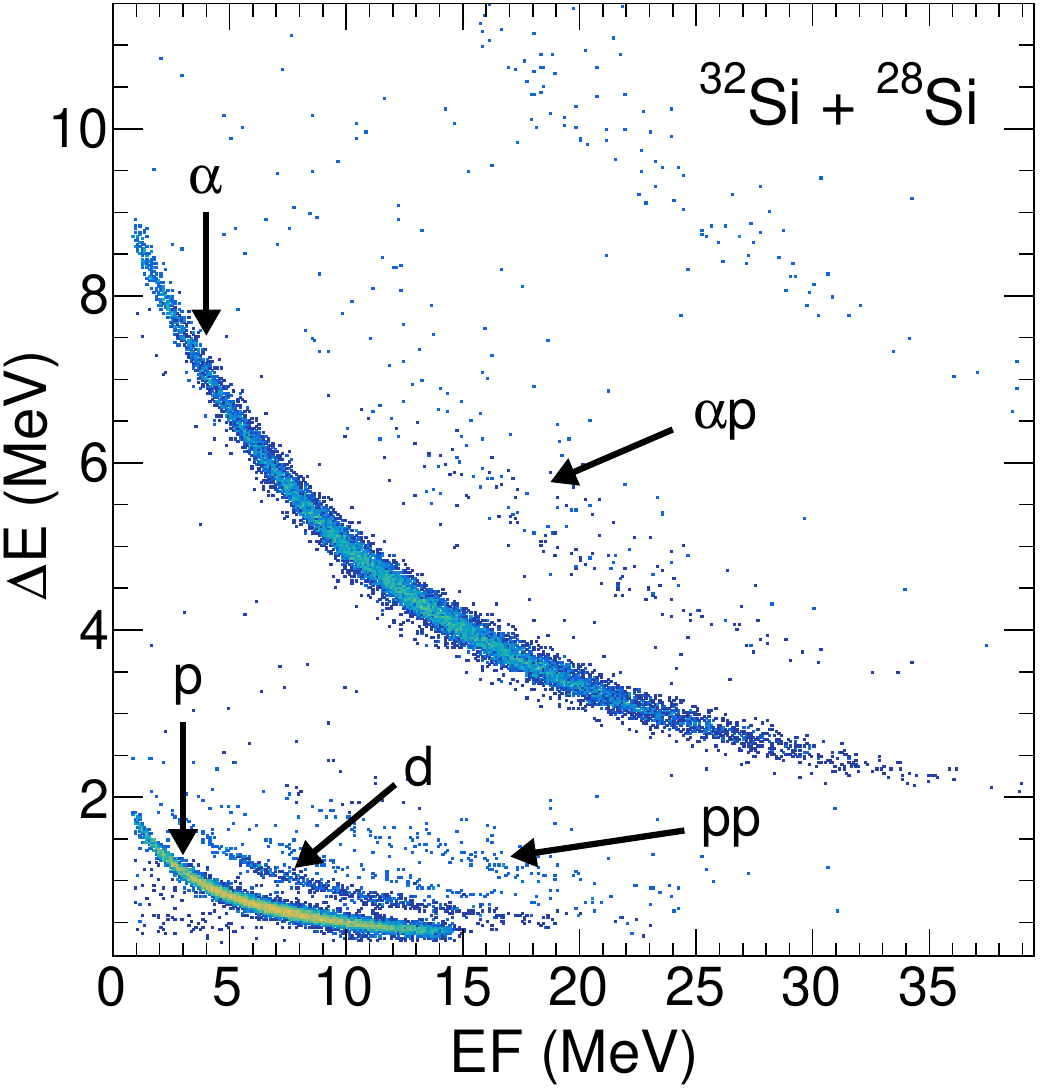}
\caption{A typical PID spectrum for light-charged particles detected in a single HiRA telescope.
}
\label{fig:HiRA2D}
\end{center}
\end{figure}

\begin{figure}
\begin{center}
\includegraphics[scale=0.43]{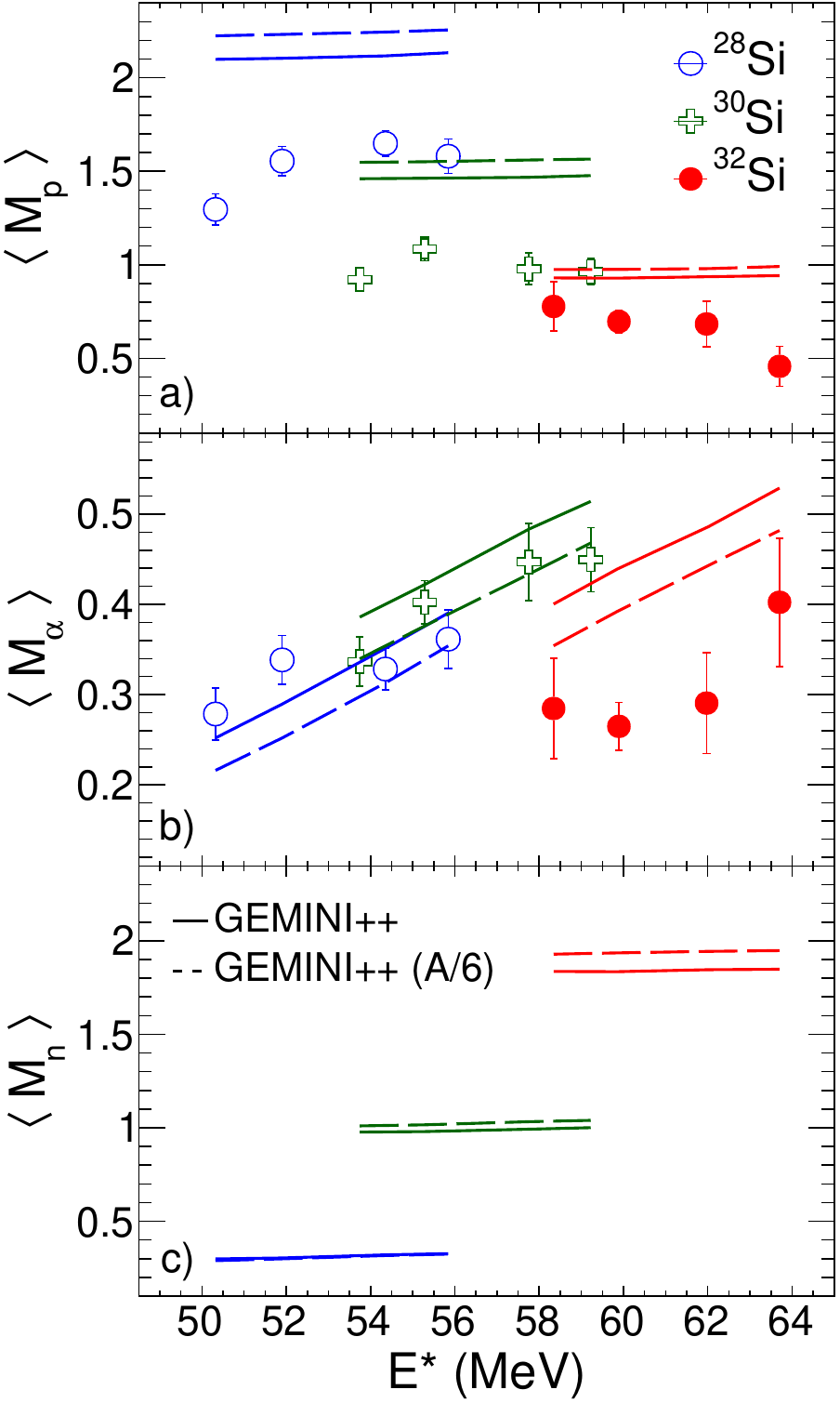}
\caption{Dependence of the multiplicity of protons and $\alpha$-particles on excitation energy. Error bars indicate the statistical errors. The GEMINI++ calculations are shown both for the default value of the level density parameter, A/7, (solid line) as well as for a value of A/6 (dashed line). Also shown for completeness, not measured in the experiment, are GEMINI++ predictions for neutrons.
}
\label{fig:Multip}
\end{center}
\end{figure}

Particle identification of LCPs with the HiRA telescopes utilized the well-established  $\Delta$E-E technique \cite{Wallace07}. A typical $\Delta$E-E spectrum for a HiRA telescope for $^{32}$Si + $^{28}$Si is shown in Fig.~\ref{fig:HiRA2D}. The dominant bands observed in the spectrum, well separated from each other, correspond to detection of protons and $\alpha$-particles. Bands situated slightly above the proton band are associated with detection of deuterons and tritons. As expected, the protons dominate the yield of hydrogen isotopes. Also evident in the spectrum are the bands associated with coincidence summing within the detector of two protons (pp) or an $\alpha$ and a proton ($\alpha$p). 

Having identified the protons and $\alpha$-particles we extract the multiplicities for these particles for all the systems measured. 
The efficiency for detection of a proton or an $\alpha$-particle in coincidence with an ER was determined using GEMINI++. Particles predicted by GEMINI++ in the center-of-mass frame were first kinematically boosted into the laboratory frame and then filtered by the detector geometric acceptance. For a particle of type $i$, the coincidence efficiency, $\epsilon_{i-ER}$, was determined from the ratio of the detected coincident particles to the total number of emitted particles. The efficiency for detecting an ER alone, $\epsilon_{ER}$, corresponded to the ratio of detected ERs to total ERs. With these two efficiencies and the total number of measured coincidences, $N_{i-ER}$, and ERs, $N_{ER}$, the multiplicities were calculated.  
\begin{equation}
    \langle M_i\rangle = \frac{N_{i-ER}}{N_{ER}}\times\frac{\epsilon_{ER}}{\epsilon_{i-ER}}
\end{equation}
The multiplicities as a function of the excitation energy, E$^*$, are shown in Fig.~\ref{fig:Multip} along with the predictions of the GEMINI++ model. The excitation energy E$^*$ reflects not only the incident energy E$_{c.m.}$ but also the fusion Q-value for each system. These Q-values are 10.922, 14.303, 18.902 MeV for $^{28}$Si, $^{30}$Si, and $^{32}$Si respectively. The quantity E$^*$ corresponds to the {\em initial} excitation of the compound system. 

For the protons one observes that the average multiplicity, $\langle M_{p}\rangle$, is approximately constant with excitation energy for a given system and decreases from $\sim$1.5 for $^{28}$Si to $\sim$0.65 for $^{32}$Si. While a significant decrease in $\langle M_{p}\rangle$ is observed from $^{28}$Si to $^{30}$Si, the decrease from $^{30}$Si to $^{32}$Si is significantly less. GEMINI++ overpredicts the $\langle M_{p}\rangle$ in all cases, though the magnitude of the overprediction decreases with increasing neutron-richness.
The measured average $\alpha$ multiplicity, $\langle M_{\alpha}\rangle$, is between $\sim$0.25-0.5 for all three systems. For $^{28,30}$Si an approximately linear dependence of $\langle M_{\alpha}\rangle$ on excitation energy is manifested. A comparable excitation energy dependence is {\em not} observed for $^{32}$Si. The $\langle M_{\alpha}\rangle$ for $^{32}$Si lies well below the expectation from an extrapolation of the linear trend observed for $^{28,30}$Si.
In the case of $^{28,30}$Si the GEMINI++ model provides a reasonable description the $\langle M_{\alpha}\rangle$ while for $^{32}$Si it significantly overpredicts the measured multiplicity.  
For completeness the average neutron multiplicity, $\langle M_{n}\rangle$ predicted by the GEMINI++ model is shown in Fig.~\ref{fig:Multip}c. The neutron multiplicity is independent of the excitation energy for a given system in the energy interval shown as observed for the proton multiplicity. As might be expected for increased neutron-excess, the $\langle M_{n}\rangle$ increases from $\sim$0.3 for $^{28}$Si to $\sim$1.9 for $^{32}$Si.

\begin{figure}
\begin{center}
\includegraphics[scale=0.45]{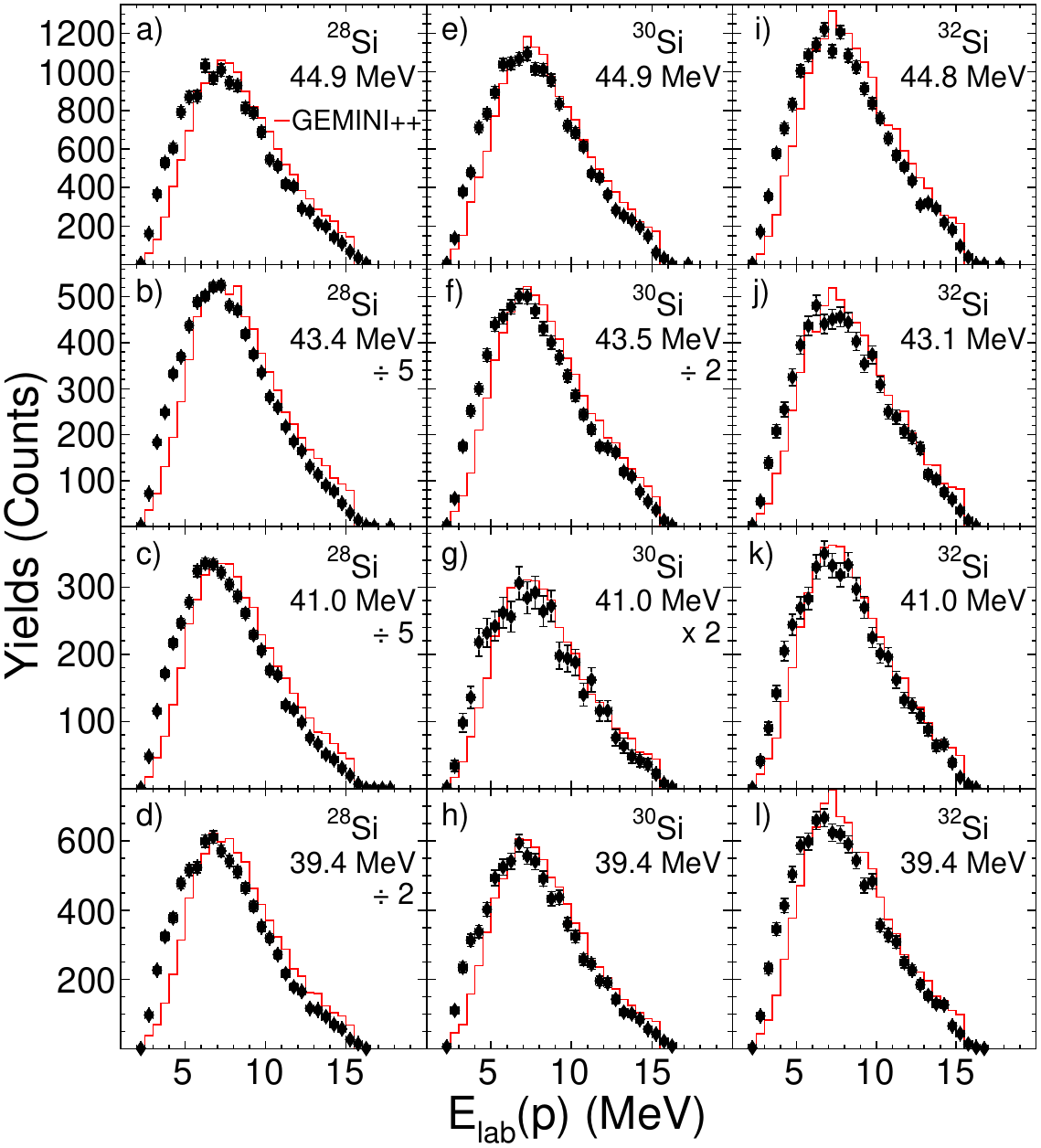}
\caption{Energy distributions of the protons in laboratory frame for different incident energies for the three systems. The solid (red) line represents the GEMINI++ prediction using the default level density parameter (A/7).
}
\label{fig:ProtonE9}
\end{center}
\end{figure}

\begin{figure}
\begin{center}
\includegraphics[scale=0.45]{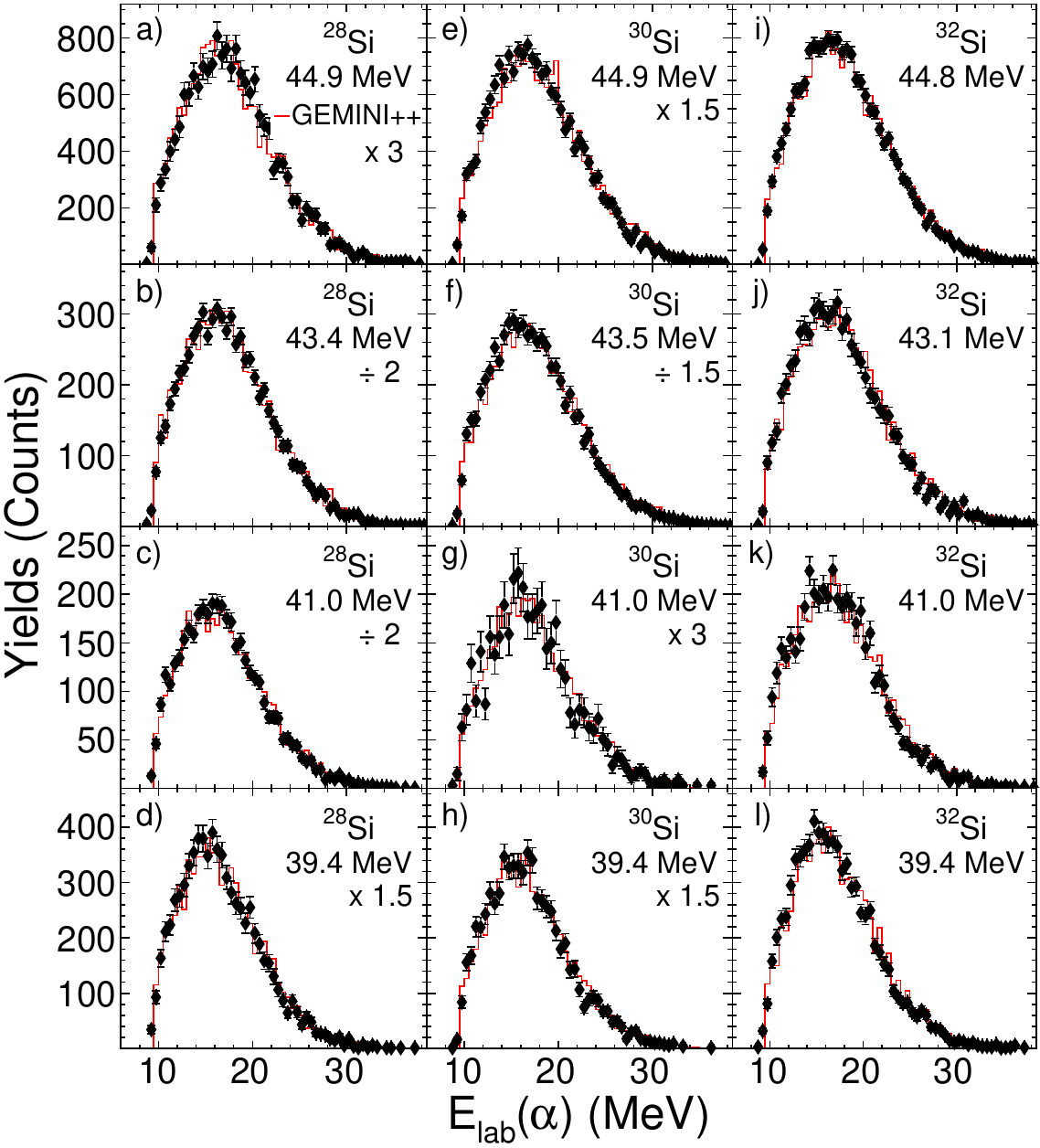}
\caption{Energy distributions of $\alpha$ particles in the laboratory frame for different incident energies for the three systems. The GEMINI++ predictions using the default level density parameter (A/7) are depicted by the solid (red) lines.
}
\label{fig:AlphaE9}
\end{center}
\end{figure}

Presented in Fig.~\ref{fig:ProtonE9} are the energy distributions of the detected protons at different incident energies for the three systems examined. These distributions are compared with the energy distributions predicted by the GEMINI++ model indicated by the solid histogram (red). To facilitate the comparison, each GEMINI++ distribution has been normalized to the integral of the experimental distribution. For all energies examined, a reasonably good overall description of the data by the model is observed. Close examination of these energy distributions however  reveals that the model underpredicts the yield at low energies while providing a better description of above-barrier energies. This relative under-prediction of low-energy protons exists for all three systems examined at all incident energies. A hint of an overprediction at higher energies is observed.

Presented in Fig.~\ref{fig:AlphaE9} are the energy distributions for the detected $\alpha$-particles along with the energy distribution predicted by the GEMINI++ model. As with the protons, the GEMINI++ energy distribution for $\alpha$-particles has been normalized to the integral of the experimental distribution. In contrast with the proton case however, the $\alpha$-particle energy distributions predicted by GEMINI++ are in very good agreement with the experimental ones.

\begin{figure}
\begin{center}
\includegraphics[scale=0.40]{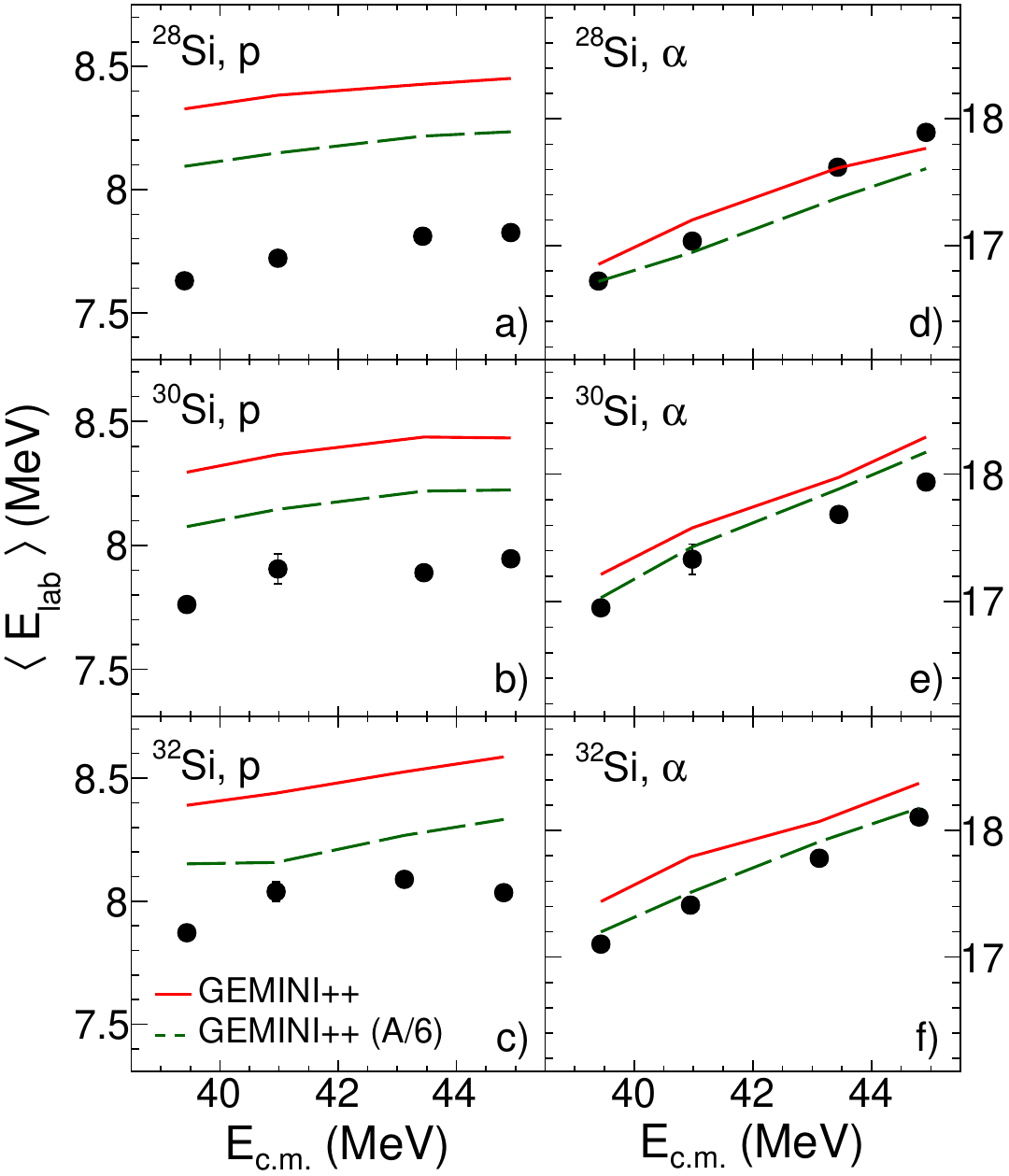}
\caption{Dependence of the average energy for protons and $\alpha$-particles on incident energy for $^{28,30,32}$Si. Error bars, in most cases the smaller than the symbol size, indicate the statistical uncertainty. The predictions of the GEMINI++ statistical model code with the level density parameter of A/7 (default) and A/6 are indicated by the solid (red) and dashed (green) lines.
}
\label{fig:Mean_pa_E}
\end{center}
\end{figure}

In Fig.~\ref{fig:Mean_pa_E} a quantitative comparison is made of the dependence of the average energy, $\langle$E$_{lab}$$\rangle$, on incident energy for both protons and $\alpha$-particles. An approximately linear dependence of the average energy of the emitted particle on E$_{c.m.}$ is observed. This linear dependence is stronger for $\alpha$-particles as compared to protons. This stronger dependence can be understood as due to the presence of a larger barrier for $\alpha$-emission as compared to proton emission. The average energy predicted by the GEMINI++ model is also presented. It is noteworthy that the slope of the average energy of the emitted particle with incident energy is reasonably well reproduced by the model for both protons and $\alpha$-particles. As might be expected from Fig.~\ref{fig:ProtonE9} for protons the model overpredicts the experimental data for all systems but the magnitude of the overprediction decreases with increasing neutron-richness. In the case of $\alpha$-particles the model is in quite good agreement with the experimental data for all three systems as expected from Fig.~\ref{fig:AlphaE9}.

\begin{figure}
\begin{center}
\includegraphics[scale=0.43]{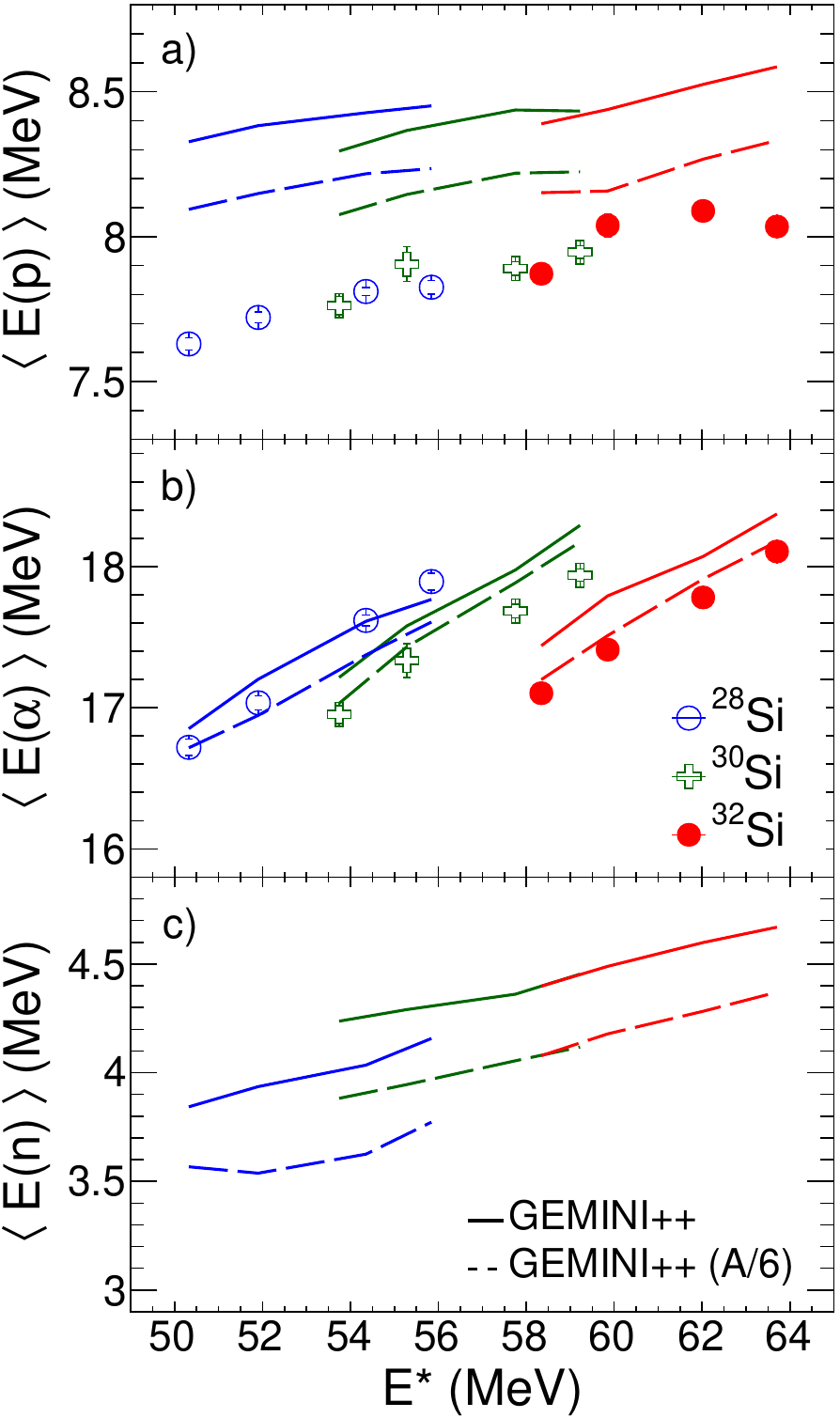}
\caption{Dependence of $\langle$E$_{lab}$$\rangle$ for neutrons, protons, and $\alpha$-particles on excitation energy, E$^*$, for all three systems. GEMINI++ predictions are shown by solid (red) and dahsed (green) lines.
}
\label{fig:MeanE_pad_estar}
\end{center}
\end{figure}

The dependence of $\langle$E$_{lab}$$\rangle$ on excitation energy, E$^*$ for the different systems is examined in 
Fig.~\ref{fig:MeanE_pad_estar}. In panel a) one observes that protons for all three systems fall on a common line. 
This result indicates that neutron-richness of the compound system does not alter to any significant extent the average kinetic energy of the emitted proton. In marked contrast the $\langle$E$_{lab}$$\rangle$ for $\alpha$-particles shown in panel b) exhibits a definite system dependence. 
The relationship predicted between the $\langle$E$_{lab}$$\rangle$ of the emitted particle and excitation energy by the statistical model code GEMINI++ is shown by the solid lines in Fig.~\ref{fig:MeanE_pad_estar}. The model predicts, as observed in the experimental data, that for protons a near system independence is observed while $\alpha$-emission manifests a system dependence. 
Close examination of the model predictions in Fig.~\ref{fig:MeanE_pad_estar}a shows that while $\langle$E$_{lab}$$\rangle$ for $^{30,32}$Si follow a common line, $\langle$E$_{lab}$$\rangle$ for $^{28}$Si is slightly larger than the value expected from this common line. In Fig.~\ref{fig:MeanE_pad_estar}c one observes that for neutrons while $\langle$E$_{lab}$$\rangle$ for $^{30,32}$Si follow a common line, $\langle$E$_{lab}$$\rangle$ for $^{28}$Si is slightly smaller than the value expected from this common line. This displacement in $\langle$E$_{lab}$$\rangle$ for $^{28}$Si as compared to $^{30,32}$Si depends sensitively on the order in which the particles are emitted as subsequently explained.

As a simple mass number dependent level-density parameter might not be accurate with increasing neutron-richness, we explored the impact of using a level density parameter of A/6 \cite{Behkami02}. The result of this increased level density parameter is represented by a dashed line in Fig.~\ref{fig:Multip}, Fig.~\ref{fig:Mean_pa_E}, and Fig.~\ref{fig:MeanE_pad_estar}. Close examination of Fig.~\ref{fig:MeanE_pad_estar} suggests that the level density parameter might change from the default value of A/7 to A/6 as one goes from $^{28}$Si to $^{32.}$Si. While this results in lowering the $\langle$E$_{lab}$$\rangle$ and $\langle M_{\alpha}\rangle$, it does not resolve the overprediction  of the proton $\langle$E$_{lab}$$\rangle$ or the overprediction of the $\langle M_{p}\rangle$, indeed it makes the discrepancy in the proton multiplicity slightly worse. 
The energy distributions predicted by GEMINI++ with level density parameter of A/6 differ only slightly from those shown in Fig.~\ref{fig:ProtonE9} and Fig.~\ref{fig:AlphaE9} as suggested by the small change in $\langle$E$_{lab}$$\rangle$.
The influence of the Yrast line on the particle multiplicities and energies was also investigated using GEMINI++. It was not possible to achieve an improved description of the experimental results by modifying the Yrast line.

\section{Discussion}

\begin{figure}
\begin{center}
\includegraphics[scale=0.43]{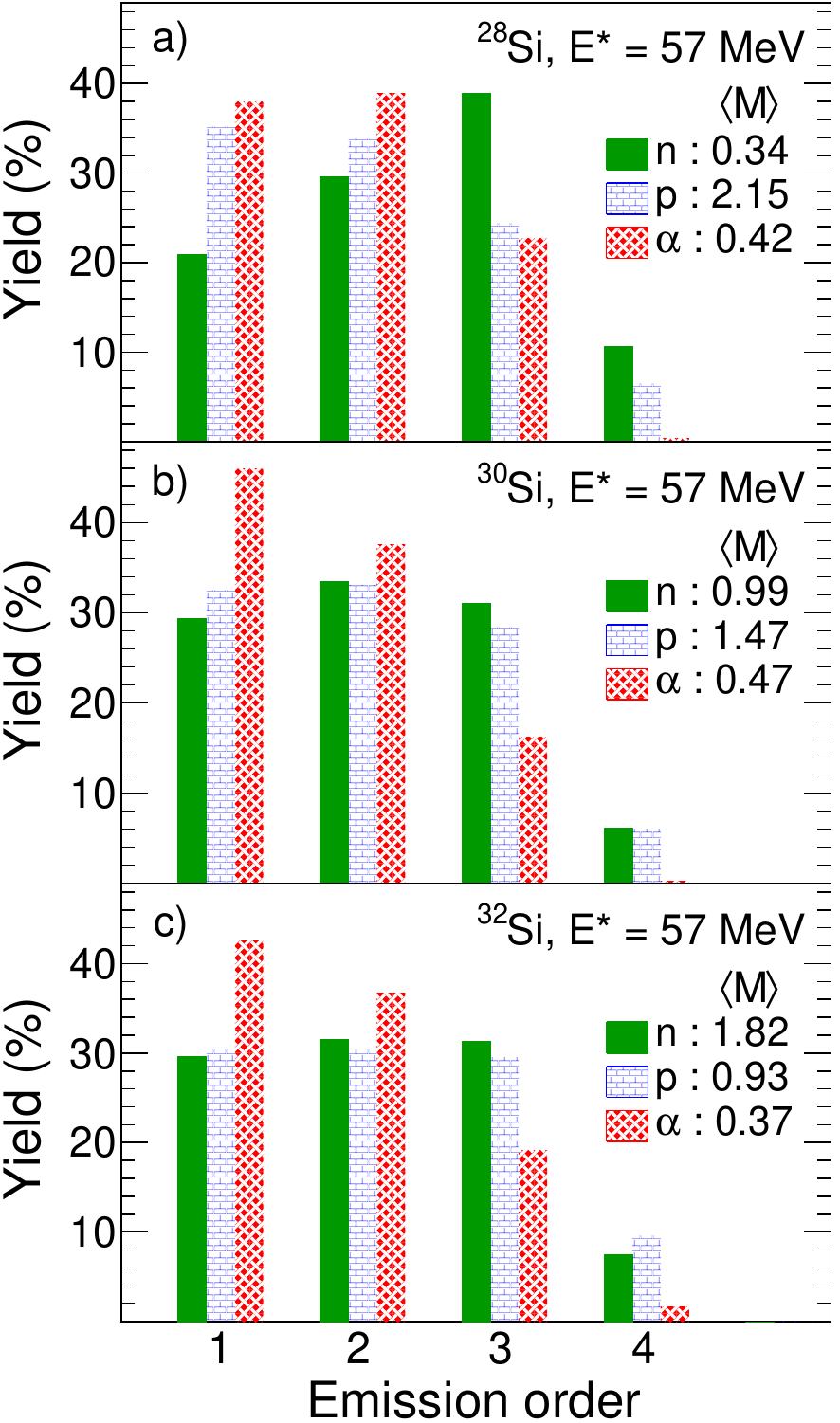}
\caption{Emission sequence of the emitted particles predicted by GEMINI++ for all three systems at E$^{*}$ = 57 MeV with a default level density parameter of A/7. 
}
\label{fig:GEMINI0}
\end{center}
\end{figure}

\begin{figure}
\begin{center}
\includegraphics[scale=0.42]{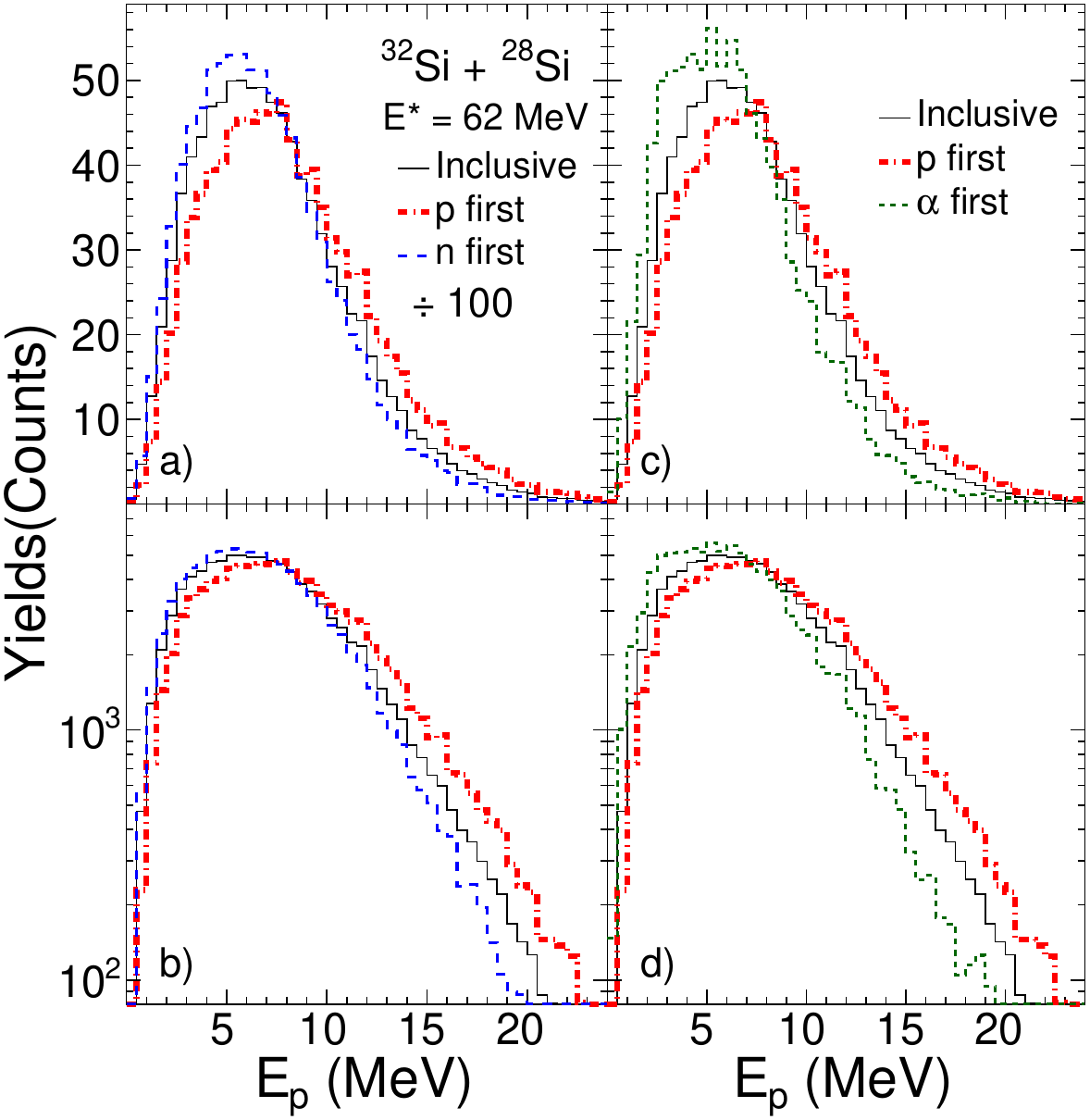}
\caption{Energy distributions of protons predicted by GEMINI++ with a default level density parameter of A/7 for the statistical decay of the $^{60}$Ni compound nucleus at E$^*$=62.0 MeV. In addition to the inclusive distribution, the distributions selected on the initial emission of a neutron, proton, or $\alpha$-particle are also presented. 
}
\label{fig:GEMINI1}
\end{center}
\end{figure}

\begin{figure}
\begin{center}
\includegraphics[scale=0.42]{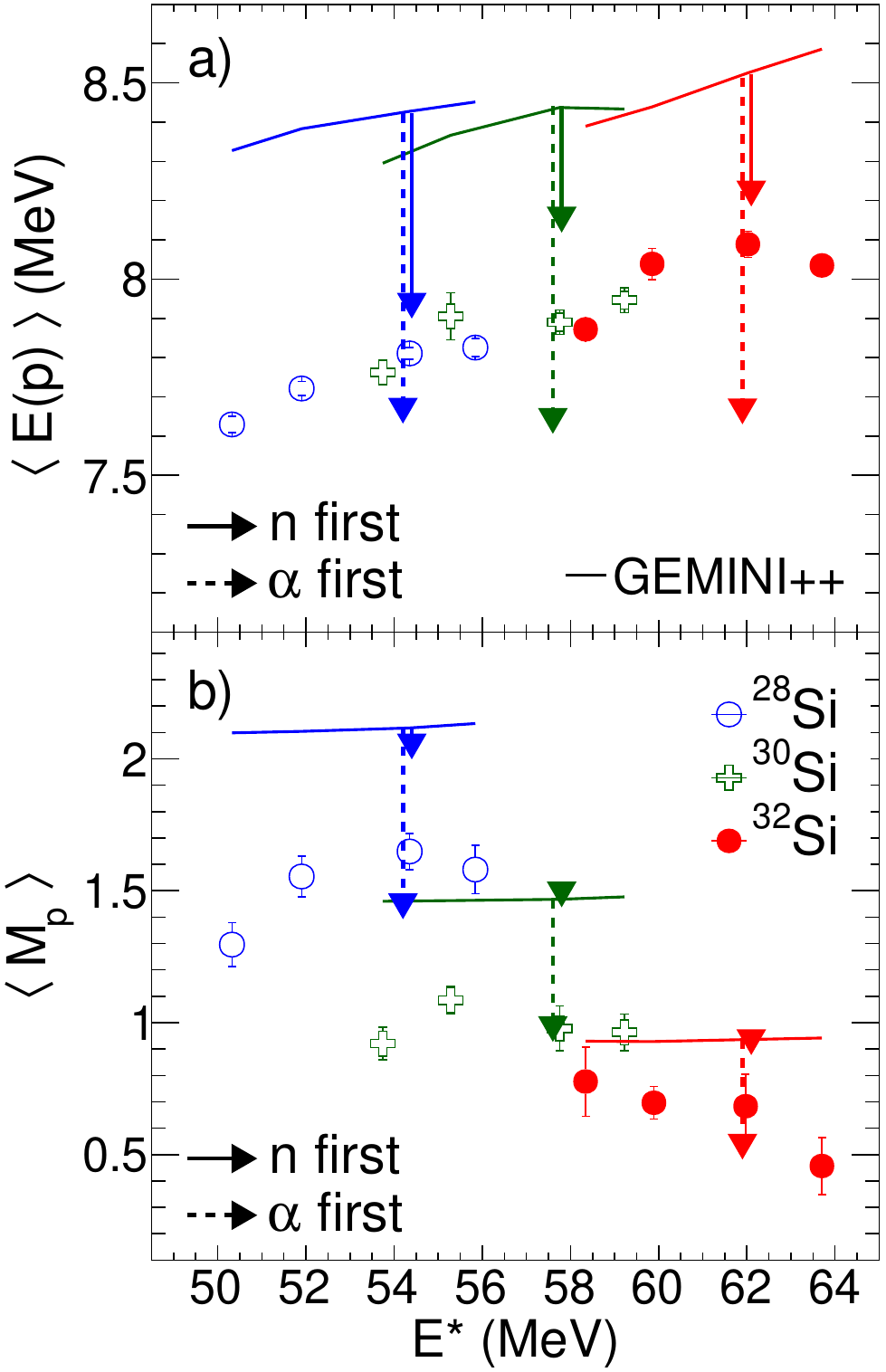}
\caption{ Comparison of the measured proton average energies and proton multiplicities as a function of excitation energy with the predictions of the GEMINI++ statistical decay code. In addition to the predictions of the default calculations (solid lines) the impact of requiring a neutron, proton, or $\alpha$-particle to be the first particle emitted is indicated by the arrows.
}
\label{fig:GEMINI2}
\end{center}
\end{figure}

\begin{figure}
\begin{center}
\includegraphics[scale=0.42]{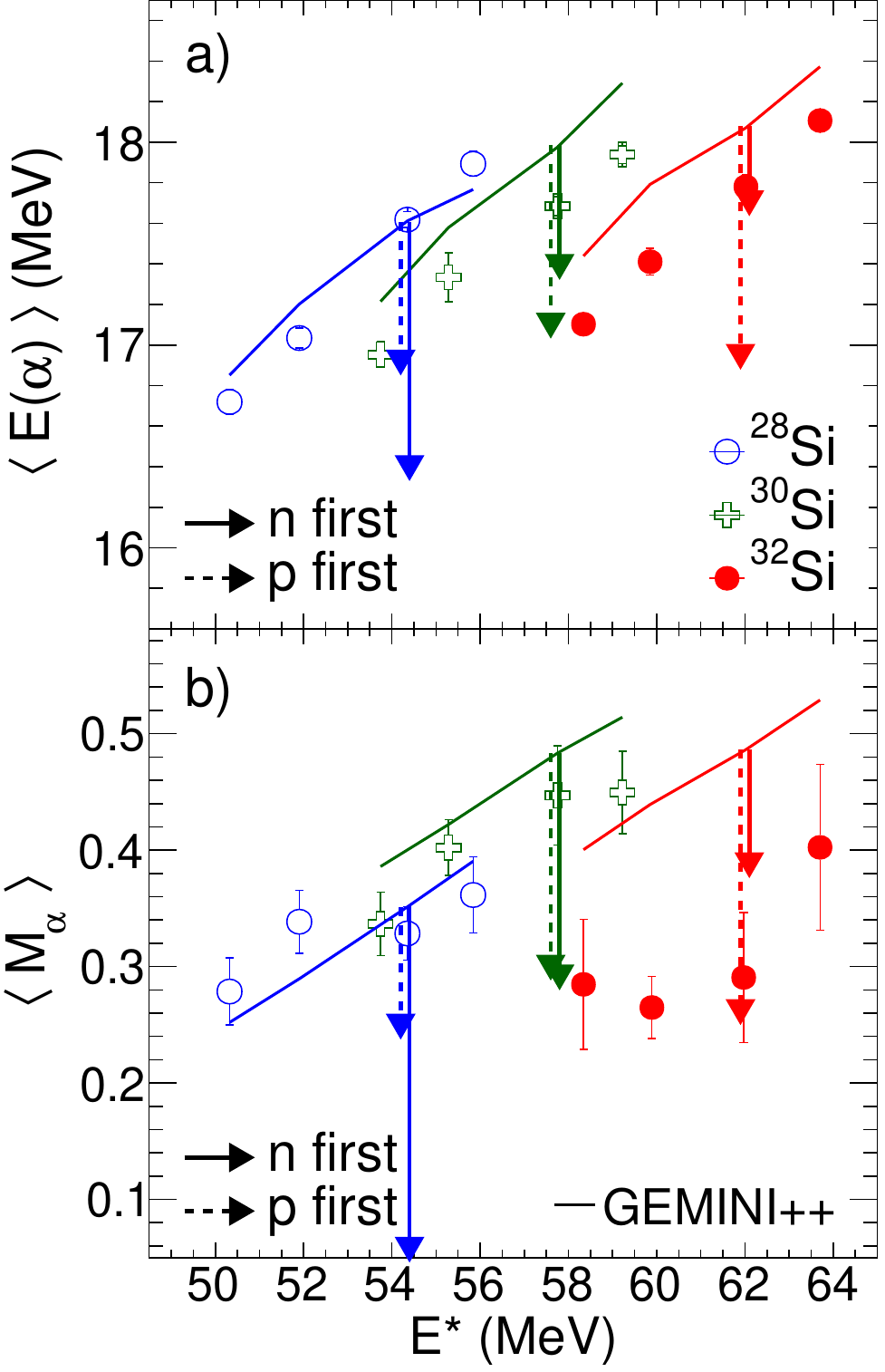}
\caption{Dependence of the average energy and multiplicity of $\alpha$-particles on excitation energy along with the GEMINI++ calculations using the default level density parameter (solid line). The influence of requiring first emission of neutron, proton, or $\alpha$-particle is indicated by the arrows. 
}
\label{fig:GEMINI3}
\end{center}
\end{figure}

Comparison of the GEMINI++ model predictions with the experimental multiplicities and energies reveals that the model provides a reasonable description of the $\alpha$-particles for $^{28,30}$Si. For $^{32}$Si while $\langle$E$_{lab}$$\rangle$ for $\alpha$-particles is reasonably described, the $\alpha$-particle multiplicity is overpredicted. Moreover, GEMINI++ systematically overpredicts both the proton multiplicities as well as their average energies for all the systems studied. This suggests a common underlying cause in the GEMINI++ calculations, largely independent of the structure differences between the systems. 

Presented in Fig.~\ref{fig:GEMINI0} is the emission sequence of protons, neutrons, and $\alpha$-particles predicted by GEMINI++. For each system, corresponding to compound nuclei of $^{56,58,60}$Ni, decay from an initial excitation energy, E$^*$=57 MeV was calculated with a default level density parameter of a=A/7. The distribution for each particle-type has been independently normalized allowing examination of their relative emission in the de-excitation cascade. The average multiplicity for each emitted particle is also indicated in Fig.~\ref{fig:GEMINI0}. For $^{28}$Si, one observes that protons and $\alpha$-particles have about the same fractional yield at each of the first three steps in the emission/de-excitation sequence. Neutrons in contrast are on average emitted later in the de-excitation cascade. In contrast, for both $^{30}$Si and $^{32}$Si, the fraction of neutrons to protons is approximately constant for all steps in the de-excitation cascade. It is interesting to note that independent of the neutron-richness of the system the fractional yield of $\alpha$-particles emitted in the first step remains approximately constant at $\sim$40$\%$. This result is likely due to the higher emission barrier for $\alpha$-particles as compared to protons which favors their early emission before the compound nucleus has de-excited through emission of other particles. The impact of increased neutron-richness is evident in the cases of $^{30,32}$Si by the enhanced early emission of neutrons as compared to the case of $^{28}$Si. This trend is qualitatively consistent, as expected, with the decreasing neutron separation energies of 
16.64 MeV, 12.22 MeV, and 11.38 MeV for $^{56}$Ni, $^{58}$Ni, and $^{60}$Ni respectively. It is this difference in the emission sequence for $^{28}$Si as compared to $^{30,32}$Si that is responsible for the difference in the $\langle$E$_{lab}$$\rangle$ observed in Fig.~\ref{fig:MeanE_pad_estar} for the GEMINI++ calculations.

The energy spectrum of an emitted particle is a complex quantity that reflects the competition of all emitted particles between one another at each step of the de-excitation cascade. The sensitivity of the energy spectra to the emission sequence was explored using the GEMINI++ model. The cases considered are not intended to reproduce the experimentally measured observables but simply to investigate the consequence of changes in the emission sequence. As the intent is to directly examine the impact of the particle emission order within GEMINI++ on the proton energy spectrum, the results presented in Fig.~\ref{fig:GEMINI1} are not filtered by the experimental acceptance. The impact of changing the balance between proton, neutron, or $\alpha$-particle as the first particle emitted is examined and the results are presented in both linear and log scale.
For a given system and excitation, GEMINI++ calculations were performed with the default level density parameter of A/7 and the resulting de-excitation cascades were selected based upon emission of a particular particle in the first step.

In Fig.~\ref{fig:GEMINI1}a the inclusive proton energy spectrum as well as with selection on initial emission of a neutron or proton is shown. The inclusive case corresponds to the GEMINII++ predictions without any artificial constraint. As qualitatively expected, selection of de-excitation events/cascades in which initial emission of a neutron occurs enhances the low-energy portion of the proton energy spectrum as it reduces the excitation at which the subsequent proton is emitted. 
The associated depletion of high energy yield is clearly evident in Fig.~\ref{fig:GEMINI1}b where the steeper slope of the energy spectrum is observed when a neutron is emitted first. In Fig.~\ref{fig:GEMINI1}c and Fig.~\ref{fig:GEMINI1}d the impact of requiring initial emission of an $\alpha$-particle is examined. 
As in the case of initial neutron emission, initial emission of an $\alpha$-particle results in the enhancement of low-energy protons at the expense of high energy protons. Initial $\alpha$ emission both lowers the excitation and reduces the barrier for subsequent proton emission. Initial $\alpha$-particle emission results in a larger increase in the yield of low-energy protons than initial emission of a neutron. 
For reference, in each case, the consequence of the initial emission being a proton is also presented. As might be expected, initial emission of a proton preferentially populates the higher energy part of the spectrum. These calculations suggest that the experimentally observed enhancement of protons at low energy as compared to the model reflects the fact that the initial emission of neutrons and/or $\alpha$-particles is under-represented in GEMINI++ as compared to the initial emission of protons.

To more quantitatively examine the impact of the choice of initial emission on the $\langle$E$_{lab}$$\rangle$ for protons, the selected GEMINI++ decays were kinematically boosted, filtered by the detector acceptance and the $\langle$E$_{lab}$$\rangle$ was calculated. The resulting $\langle$E$_{lab}$$\rangle$ are compared with the experimental data in Fig.~\ref{fig:GEMINI2}a. One observes that requiring that the first particle emitted be either a neutron or an $\alpha$-particle lowers $\langle$E$_{lab}$$\rangle$ for protons. First emission of a neutron (solid arrows) results in improved agreement with the experimental data, though with some overprediction. Requiring first emission of an $\alpha$-particle (dashed arrows) results in too low of an $\langle$E$_{lab}$$\rangle$, particularly for $^{30,32}$Si.

Presented in Fig.~\ref{fig:GEMINI2}b is the impact of requiring first emission of a neutron or $\alpha$-particle on the average proton multiplicity,
$\langle$M$_p$$\rangle$. Evident in Fig.~\ref{fig:GEMINI2}b is that initial emission of a neutron does not effectively alter the average proton multiplicity while the initial emission of the
$\alpha$-particle does bring the predicted proton multiplicities into good agreement with the data. 
These results suggest that enhanced initial $\alpha$-particle emission in the model is necessary to achieve a better overall description of the experimental data.
While $\langle$E$_{lab}$$\rangle$ for protons and $\langle$M$_{p}$$\rangle$ will be improved, it should be realized that increased initial $\alpha$ emission will suppress the yield of low energy $\alpha$-particles predicted by the model, somewhat worsening the description of the experimental $\alpha$ energy distributions.

In Fig.~\ref{fig:GEMINI3} the consequence of initial emission of a neutron or proton on the $\langle$E$_{lab}$$\rangle$ of $\alpha$-particles is investigated. For both $^{28,30}$Si the default calculations provide a reasonably good description of $\langle$E$_{lab}$$\rangle$ and $\langle$M$_{\alpha}$$\rangle$ with a overprediction of both quantities for $^{32}$Si. Increased initial emission of either neutrons or protons significantly lowers $\langle$E$_{lab}$$\rangle$ and 
$\langle$M$_{\alpha}$$\rangle$ for $^{28,30}$Si well below the experimental data. As the default GEMINI++ calculations already lie close to the 
experimental data for 
$^{28,30}$Si, it is unsurprising that modifying the initial emission results in a larger discrepancy between model and data.
In the case of $^{32}$Si, on the basis $\langle$M$_{\alpha}$$\rangle$ increased initial emission of a nucleon is indicated while consideration of $\langle$E$_{lab}$$\rangle$ reveals that this particle should be a neutron. While initial emission of a neutron results in a reasonably good description for 
$\langle$E$_{lab}$$\rangle$, the predicted $\langle$M$_{\alpha}$$\rangle$ is still somewhat high suggesting that changing the sequence of particle emission alone is insufficient. The discrepancy suggests that for $^{32}$Si, nucleon emission, relative to $\alpha$ emission, might be under-represented in the model.

\section{Conclusions}

The coincident measurement of the light-charged particles (LCPs) associated with fusion in $^{28,30,32}$Si + $^{28}$Si provided a rich dataset to investigate the statistical decay of the emitted particles. Selection on the identity of the emitted particle reveals how the measured angular distribution of ERs encodes information of the balance between nucleon and $\alpha$-particle emission. Calculations with the statistical model code GEMINI++ support this de-convolution of the angular distribution with some deviations observed. Detection of the ERs in coincidence with LCPs facilitated determination of the average proton and $\alpha$-particle multiplicities associated with fusion. The measured energy distributions of protons and $\alpha$-particles were compared with the predictions of the GEMINI++ model. While the $\alpha$-particle energy distributions were well described, a systematic underprediction of the low-energy portion of the proton energy distribution was observed. Comparison of the average energy dependence on E$^*$ for protons and $\alpha$-particles with the GEMINI++ model predictions quantified this discrepancy. While the GEMINI++ statistical model provides a reasonable description of the $\alpha$-particles multiplicities and energies it overpredicts the average proton multiplicities and energies. With increasing neutron-richness the discrepancy between GEMINI++ predictions and the experimental data for protons decreases while increasing slightly for $\alpha$-particles.

Changing standard model parameters in GEMINI++, like level densities, was not sufficient to reproduce the measured proton energy distributions or multiplicities. Although increasing the level density parameter lowers the $\langle$E$_{lab}$$\rangle$, the model still over-predicts the proton average energy. Moreover, the overprediction of the proton multiplicities is increased. 

The impact of modifying the sequence of particle emissions on the average energy and multiplicity was examined. By increasing the fractional yield of $\alpha$-particle emission in the initial step, the model predictions for $\langle$E$_{lab}$$\rangle$ and $\langle$M$\rangle$ for protons can be brought into better agreement with the experimental data. For events which contain an $\alpha$-particle, requiring that an $\alpha$-particles is emitted first in the de-excitation sequence does not significantly alter the $\alpha$ multiplicity. It does however slightly increase the $\alpha$-particle energy resulting in on average a higher energy $\alpha$-particles while depleting the yield of lower energy $\alpha$-particles. 
This increased yield of higher energy $\alpha$-particles will provide a larger transverse momentum to the ER, resulting in its observation at larger angles. 
Consequently, the underprediction of the ER angular distribution for $^{28,30}$Si by the model in the angular range $\theta_{lab}$$>$12$^\circ$,  will be reduced. 
For the case of $^{32}$Si, increased initial neutron emission as compared to the model calculations is also indicated.

This work reveals that joint high-quality measurement of the particle energy distributions as well as their multiplicities, together with the angular distributions of ERs, can provide information on the details of the particle emission sequence. Measurement of the neutron kinetic energies and multiplicity, though challenging, would be invaluable in further constraining the emission sequence.

\begin{acknowledgments}
We acknowledge the high quality beams provided by the staff at NSCL, Michigan Sate University that made this experiment possible. We are thankful for the high-quality services of the Mechanical Instrument Services and Electronic Instrument Services facilities at Indiana University. This work was supported by the U.S. Department of Energy Office of Science under Grant Nos. DE-FG02-88ER-40404 (Indiana University), 
DE-SC0025230 (Indiana University), DE-SC0021938, DE-SC0022299 (Michigan State University), and the National Science Foundation under PHY-1712832. A.C. gratefully acknowledges the support from GANIL, CNRS-IN2P3 and CEA-DRF. Support from the Australian Research Council is gratefully acknowledged by KJC (DE230100197). KWB is supported in part by the National Science Foundation under Grant No. PHY-2309923 and U.S. Department of Energy, Office of Science, Nuclear Physics under Award No. DE-SC0021938.

\end{acknowledgments}
\appendix

%\bibliography{Si_Si}% Produces the bibliography via BibTeX.

\begin{thebibliography}{25}
\expandafter\ifx\csname natexlab\endcsname\relax\def\natexlab#1{#1}\fi
\expandafter\ifx\csname bibnamefont\endcsname\relax
  \def\bibnamefont#1{#1}\fi
\expandafter\ifx\csname bibfnamefont\endcsname\relax
  \def\bibfnamefont#1{#1}\fi
\expandafter\ifx\csname citenamefont\endcsname\relax
  \def\citenamefont#1{#1}\fi
\expandafter\ifx\csname url\endcsname\relax
  \def\url#1{\texttt{#1}}\fi
\expandafter\ifx\csname urlprefix\endcsname\relax\def\urlprefix{URL }\fi
\providecommand{\bibinfo}[2]{#2}
\providecommand{\eprint}[2][]{\url{#2}}

\bibitem[{\citenamefont{Rose and Jones}(1984)}]{Rose84}
\bibinfo{author}{\bibfnamefont{H.~J.} \bibnamefont{Rose}} \bibnamefont{and}
  \bibinfo{author}{\bibfnamefont{G.~A.} \bibnamefont{Jones}},
  \bibinfo{journal}{Nature} \textbf{\bibinfo{volume}{307}},
  \bibinfo{pages}{245} (\bibinfo{year}{1984}).

\bibitem[{\citenamefont{Horiuchi}(2010)}]{Horiuchi10}
\bibinfo{author}{\bibfnamefont{H.}~\bibnamefont{Horiuchi}},
  \emph{\bibinfo{title}{Clusters in Nuclei vol. 818}}
  (\bibinfo{publisher}{Springer}, \bibinfo{year}{2010}).

\bibitem[{\citenamefont{Hoyle}(1954)}]{Hoyle54}
\bibinfo{author}{\bibfnamefont{F.}~\bibnamefont{Hoyle}},
  \bibinfo{journal}{Astrophys. J. Suppl.195} \textbf{\bibinfo{volume}{1}},
  \bibinfo{pages}{121} (\bibinfo{year}{1954}).

\bibitem[{\citenamefont{Rose et~al.}(1957)\citenamefont{Rose, Fowler,
  Lauritsen, and Lauritsen}}]{Cook57}
\bibinfo{author}{\bibfnamefont{H.~J.} \bibnamefont{Rose}},
  \bibinfo{author}{\bibfnamefont{W.~A.} \bibnamefont{Fowler}},
  \bibinfo{author}{\bibfnamefont{C.~C.} \bibnamefont{Lauritsen}},
  \bibnamefont{and}
  \bibinfo{author}{\bibfnamefont{T.}~\bibnamefont{Lauritsen}},
  \bibinfo{journal}{Phys. Rev.} \textbf{\bibinfo{volume}{107}},
  \bibinfo{pages}{508} (\bibinfo{year}{1957}).

\bibitem[{\citenamefont{Neff and Feldmeier}(2008)}]{Neff08}
\bibinfo{author}{\bibfnamefont{T.}~\bibnamefont{Neff}} \bibnamefont{and}
  \bibinfo{author}{\bibfnamefont{H.}~\bibnamefont{Feldmeier}},
  \bibinfo{journal}{Int. J. Mod. Phys. E} \textbf{\bibinfo{volume}{17}},
  \bibinfo{pages}{2005} (\bibinfo{year}{2008}).

\bibitem[{\citenamefont{Röpke et~al.}(2013)\citenamefont{Röpke, Bastian,
  Blaschke, Klähn, Typel, and Wolter}}]{Ropke13}
\bibinfo{author}{\bibfnamefont{G.}~\bibnamefont{Röpke}},
  \bibinfo{author}{\bibfnamefont{N.-U.} \bibnamefont{Bastian}},
  \bibinfo{author}{\bibfnamefont{D.}~\bibnamefont{Blaschke}},
  \bibinfo{author}{\bibfnamefont{T.}~\bibnamefont{Klähn}},
  \bibinfo{author}{\bibfnamefont{S.}~\bibnamefont{Typel}}, \bibnamefont{and}
  \bibinfo{author}{\bibfnamefont{H.~H.} \bibnamefont{Wolter}},
  \bibinfo{journal}{Nucl. Phys. A} \textbf{\bibinfo{volume}{897}},
  \bibinfo{pages}{70} (\bibinfo{year}{2013}).

\bibitem[{\citenamefont{Typel et~al.}(2010)\citenamefont{Typel, Röpke, Klähn,
  Blaschke, and Wolter}}]{Typel10}
\bibinfo{author}{\bibfnamefont{S.}~\bibnamefont{Typel}},
  \bibinfo{author}{\bibfnamefont{G.}~\bibnamefont{Röpke}},
  \bibinfo{author}{\bibfnamefont{T.}~\bibnamefont{Klähn}},
  \bibinfo{author}{\bibfnamefont{D.}~\bibnamefont{Blaschke}}, \bibnamefont{and}
  \bibinfo{author}{\bibfnamefont{H.~H.} \bibnamefont{Wolter}},
  \bibinfo{journal}{Phys. Rev C} \textbf{\bibinfo{volume}{81}},
  \bibinfo{pages}{015803} (\bibinfo{year}{2010}).

\bibitem[{\citenamefont{Avila et~al.}(2014)\citenamefont{Avila, Rogachev,
  Goldberg, Johnson, Kemper, Tchuvilsky, and Volya}}]{Avila14}
\bibinfo{author}{\bibfnamefont{M.~L.} \bibnamefont{Avila}},
  \bibinfo{author}{\bibfnamefont{G.~V.} \bibnamefont{Rogachev}},
  \bibinfo{author}{\bibfnamefont{V.~Z.} \bibnamefont{Goldberg}},
  \bibinfo{author}{\bibfnamefont{E.~D.} \bibnamefont{Johnson}},
  \bibinfo{author}{\bibfnamefont{K.~W.} \bibnamefont{Kemper}},
  \bibinfo{author}{\bibfnamefont{Y.~M.} \bibnamefont{Tchuvilsky}},
  \bibnamefont{and} \bibinfo{author}{\bibfnamefont{A.~S.} \bibnamefont{Volya}},
  \bibinfo{journal}{Phys. Rev. C} \textbf{\bibinfo{volume}{90}},
  \bibinfo{pages}{024327} (\bibinfo{year}{2014}).

\bibitem[{\citenamefont{Vadas et~al.}(2015)\citenamefont{Vadas, Steinbach,
  Schmidt, Singh, Haycraft, Hudan, deSouza, Baby, Kuvin, and
  Wiedenh\"over}}]{Vadas15}
\bibinfo{author}{\bibfnamefont{J.}~\bibnamefont{Vadas}},
  \bibinfo{author}{\bibfnamefont{T.~K.} \bibnamefont{Steinbach}},
  \bibinfo{author}{\bibfnamefont{J.}~\bibnamefont{Schmidt}},
  \bibinfo{author}{\bibfnamefont{V.}~\bibnamefont{Singh}},
  \bibinfo{author}{\bibfnamefont{C.}~\bibnamefont{Haycraft}},
  \bibinfo{author}{\bibfnamefont{S.}~\bibnamefont{Hudan}},
  \bibinfo{author}{\bibfnamefont{R.}~\bibnamefont{deSouza}},
  \bibinfo{author}{\bibfnamefont{L.~T.} \bibnamefont{Baby}},
  \bibinfo{author}{\bibfnamefont{S.~A.} \bibnamefont{Kuvin}}, \bibnamefont{and}
  \bibinfo{author}{\bibfnamefont{I.}~\bibnamefont{Wiedenh\"over}},
  \bibinfo{journal}{Phys. Rev. C} \textbf{\bibinfo{volume}{92}},
  \bibinfo{pages}{064610} (\bibinfo{year}{2015}).

\bibitem[{\citenamefont{Gavron}(1980)}]{PACE4}
\bibinfo{author}{\bibfnamefont{A.}~\bibnamefont{Gavron}},
  \bibinfo{journal}{Phys. Rev C} \textbf{\bibinfo{volume}{21}},
  \bibinfo{pages}{230} (\bibinfo{year}{1980}).

\bibitem[{\citenamefont{Tabor et~al.}(1977)\citenamefont{Tabor, Eisen, Kovar,
  and Vager}}]{Tabor77}
\bibinfo{author}{\bibfnamefont{S.~L.} \bibnamefont{Tabor}},
  \bibinfo{author}{\bibfnamefont{Y.}~\bibnamefont{Eisen}},
  \bibinfo{author}{\bibfnamefont{D.~G.} \bibnamefont{Kovar}}, \bibnamefont{and}
  \bibinfo{author}{\bibfnamefont{Z.}~\bibnamefont{Vager}},
  \bibinfo{journal}{Phys. Rev. C} \textbf{\bibinfo{volume}{16}},
  \bibinfo{pages}{673} (\bibinfo{year}{1977}).

\bibitem[{\citenamefont{Papadopoulos et~al.}(1986)\citenamefont{Papadopoulos,
  Vlastou, Gazis, Assimakopoulos, Kalfas, Kossionides, and Xenoulis}}]{Papa86}
\bibinfo{author}{\bibfnamefont{C.~T.} \bibnamefont{Papadopoulos}},
  \bibinfo{author}{\bibfnamefont{R.}~\bibnamefont{Vlastou}},
  \bibinfo{author}{\bibfnamefont{E.~N.} \bibnamefont{Gazis}},
  \bibinfo{author}{\bibfnamefont{P.~A.} \bibnamefont{Assimakopoulos}},
  \bibinfo{author}{\bibfnamefont{C.~A.} \bibnamefont{Kalfas}},
  \bibinfo{author}{\bibfnamefont{S.}~\bibnamefont{Kossionides}},
  \bibnamefont{and} \bibinfo{author}{\bibfnamefont{A.~C.}
  \bibnamefont{Xenoulis}}, \bibinfo{journal}{Phys. Rev.C}
  \textbf{\bibinfo{volume}{34}}, \bibinfo{pages}{196} (\bibinfo{year}{1986}).

\bibitem[{\citenamefont{Schuetrumpf and Nazarewicz}(2017)}]{Schuetrumpf17}
\bibinfo{author}{\bibfnamefont{B.}~\bibnamefont{Schuetrumpf}} \bibnamefont{and}
  \bibinfo{author}{\bibfnamefont{W.}~\bibnamefont{Nazarewicz}},
  \bibinfo{journal}{Phys. Rev. C} \textbf{\bibinfo{volume}{96}},
  \bibinfo{pages}{064608} (\bibinfo{year}{2017}).

\bibitem[{\citenamefont{Villari et~al.}(2023)\citenamefont{Villari, Bollen,
  Henriques, Lapierre, Nash, Ringle, Schwarz, and Sumithrarachchi}}]{Villari23}
\bibinfo{author}{\bibfnamefont{A.}~\bibnamefont{Villari}},
  \bibinfo{author}{\bibfnamefont{G.}~\bibnamefont{Bollen}},
  \bibinfo{author}{\bibfnamefont{A.}~\bibnamefont{Henriques}},
  \bibinfo{author}{\bibfnamefont{A.}~\bibnamefont{Lapierre}},
  \bibinfo{author}{\bibfnamefont{S.}~\bibnamefont{Nash}},
  \bibinfo{author}{\bibfnamefont{R.}~\bibnamefont{Ringle}},
  \bibinfo{author}{\bibfnamefont{S.}~\bibnamefont{Schwarz}}, \bibnamefont{and}
  \bibinfo{author}{\bibfnamefont{C.}~\bibnamefont{Sumithrarachchi}},
  \bibinfo{journal}{Nucl. Instr. Meth. Phys. Res. B}
  \textbf{\bibinfo{volume}{541}}, \bibinfo{pages}{350} (\bibinfo{year}{2023}).

\bibitem[{\citenamefont{Bowman and Heffner}(1978)}]{Bowman78}
\bibinfo{author}{\bibfnamefont{J.~D.} \bibnamefont{Bowman}} \bibnamefont{and}
  \bibinfo{author}{\bibfnamefont{R.~H.} \bibnamefont{Heffner}},
  \bibinfo{journal}{Nuclear Instruments and Methods}
  \textbf{\bibinfo{volume}{148}}, \bibinfo{pages}{503} (\bibinfo{year}{1978}).

\bibitem[{\citenamefont{Steinbach et~al.}(2014)\citenamefont{Steinbach,
  Rudolph, Gosser, Brown, Floyd, Hudan, deSouza, Liang, Shapira, and
  Famiano}}]{Steinbach14}
\bibinfo{author}{\bibfnamefont{T.~K.} \bibnamefont{Steinbach}},
  \bibinfo{author}{\bibfnamefont{M.~J.} \bibnamefont{Rudolph}},
  \bibinfo{author}{\bibfnamefont{Z.~Q.} \bibnamefont{Gosser}},
  \bibinfo{author}{\bibfnamefont{K.}~\bibnamefont{Brown}},
  \bibinfo{author}{\bibfnamefont{B.}~\bibnamefont{Floyd}},
  \bibinfo{author}{\bibfnamefont{S.}~\bibnamefont{Hudan}},
  \bibinfo{author}{\bibfnamefont{R.~T.} \bibnamefont{deSouza}},
  \bibinfo{author}{\bibfnamefont{J.~F.} \bibnamefont{Liang}},
  \bibinfo{author}{\bibfnamefont{D.}~\bibnamefont{Shapira}}, \bibnamefont{and}
  \bibinfo{author}{\bibfnamefont{M.}~\bibnamefont{Famiano}},
  \bibinfo{journal}{Nucl. Instr. Meth. In Phys. Res. A}
  \textbf{\bibinfo{volume}{743}}, \bibinfo{pages}{5} (\bibinfo{year}{2014}).

\bibitem[{\citenamefont{Vadas et~al.}(2016)\citenamefont{Vadas, Singh, Visser,
  Alexander, Hudan, Huston, Wiggins, Chbihi, Famiano, Bischak
  et~al.}}]{Vadas16}
\bibinfo{author}{\bibfnamefont{J.}~\bibnamefont{Vadas}},
  \bibinfo{author}{\bibfnamefont{V.}~\bibnamefont{Singh}},
  \bibinfo{author}{\bibfnamefont{G.}~\bibnamefont{Visser}},
  \bibinfo{author}{\bibfnamefont{A.}~\bibnamefont{Alexander}},
  \bibinfo{author}{\bibfnamefont{S.}~\bibnamefont{Hudan}},
  \bibinfo{author}{\bibfnamefont{J.}~\bibnamefont{Huston}},
  \bibinfo{author}{\bibfnamefont{B.~B.} \bibnamefont{Wiggins}},
  \bibinfo{author}{\bibfnamefont{A.}~\bibnamefont{Chbihi}},
  \bibinfo{author}{\bibfnamefont{M.}~\bibnamefont{Famiano}},
  \bibinfo{author}{\bibfnamefont{M.~M.} \bibnamefont{Bischak}},
  \bibnamefont{et~al.}, \bibinfo{journal}{Nucl. Instr. Meth. Phys. Res. A}
  \textbf{\bibinfo{volume}{837}}, \bibinfo{pages}{28} (\bibinfo{year}{2016}).

\bibitem[{\citenamefont{Wallace et~al.}(2007)\citenamefont{Wallace, Famiano,
  vanGoethem, Rogers, Lynch, J.Clifford, Delaunay, Lee, Labostoy, Mocko
  et~al.}}]{Wallace07}
\bibinfo{author}{\bibfnamefont{M.~S.} \bibnamefont{Wallace}},
  \bibinfo{author}{\bibfnamefont{M.~A.} \bibnamefont{Famiano}},
  \bibinfo{author}{\bibfnamefont{M.-J.} \bibnamefont{vanGoethem}},
  \bibinfo{author}{\bibfnamefont{A.}~\bibnamefont{Rogers}},
  \bibinfo{author}{\bibfnamefont{W.~G.} \bibnamefont{Lynch}},
  \bibinfo{author}{\bibnamefont{J.Clifford}},
  \bibinfo{author}{\bibfnamefont{F.}~\bibnamefont{Delaunay}},
  \bibinfo{author}{\bibfnamefont{J.}~\bibnamefont{Lee}},
  \bibinfo{author}{\bibfnamefont{S.}~\bibnamefont{Labostoy}},
  \bibinfo{author}{\bibfnamefont{M.}~\bibnamefont{Mocko}},
  \bibnamefont{et~al.}, \bibinfo{journal}{Nucl. Instr. Meth. Phys. Res. A}
  \textbf{\bibinfo{volume}{583}}, \bibinfo{pages}{302} (\bibinfo{year}{2007}).

\bibitem[{\citenamefont{deSouza}()}]{zepto}
\bibinfo{author}{\bibfnamefont{R.~T.} \bibnamefont{deSouza}},
  \emph{\bibinfo{title}{{Z}eptosystems {I}nc}}.

\bibitem[{Mic()}]{MicronSemiconductor}
\urlprefix\url{http://www.micronsemiconductor.co.uk}.

\bibitem[{\citenamefont{deSouza et~al.}(2011)\citenamefont{deSouza, Alexander,
  Brown, Floyd, Gosser, Hudan, Poehlman, and Rudolph}}]{deSouza11}
\bibinfo{author}{\bibfnamefont{R.~T.} \bibnamefont{deSouza}},
  \bibinfo{author}{\bibfnamefont{A.}~\bibnamefont{Alexander}},
  \bibinfo{author}{\bibfnamefont{K.}~\bibnamefont{Brown}},
  \bibinfo{author}{\bibfnamefont{B.}~\bibnamefont{Floyd}},
  \bibinfo{author}{\bibfnamefont{Z.~Q.} \bibnamefont{Gosser}},
  \bibinfo{author}{\bibfnamefont{S.}~\bibnamefont{Hudan}},
  \bibinfo{author}{\bibfnamefont{J.}~\bibnamefont{Poehlman}}, \bibnamefont{and}
  \bibinfo{author}{\bibfnamefont{M.~J.} \bibnamefont{Rudolph}},
  \bibinfo{journal}{Nucl. Instr. Meth. In Phys. Res. A}
  \textbf{\bibinfo{volume}{632}}, \bibinfo{pages}{133} (\bibinfo{year}{2011}).

\bibitem[{\citenamefont{Charity}(2010)}]{Charity10}
\bibinfo{author}{\bibfnamefont{R.}~\bibnamefont{Charity}},
  \bibinfo{journal}{Phys. Rev. C} \textbf{\bibinfo{volume}{82}},
  \bibinfo{pages}{014610} (\bibinfo{year}{2010}).

\bibitem[{\citenamefont{Bass}(1974)}]{Bass74}
\bibinfo{author}{\bibfnamefont{R.}~\bibnamefont{Bass}}, \bibinfo{journal}{Nucl.
  Phys. A} \textbf{\bibinfo{volume}{231}}, \bibinfo{pages}{45}
  (\bibinfo{year}{1974}).

\bibitem[{\citenamefont{Ziegler et~al.}()\citenamefont{Ziegler, Biersack, and
  Ziegler}}]{SRIM}
\bibinfo{author}{\bibfnamefont{J.~F.} \bibnamefont{Ziegler}},
  \bibinfo{author}{\bibfnamefont{J.~P.} \bibnamefont{Biersack}},
  \bibnamefont{and} \bibinfo{author}{\bibfnamefont{M.~D.}
  \bibnamefont{Ziegler}}, \emph{\bibinfo{title}{{T}he {S}topping and {Range} of
  {I}ons in {M}atter}}, \urlprefix\url{http://www.SRIM.org}.

\bibitem[{\citenamefont{Behkami et~al.}(2002)\citenamefont{Behkami, Kargar, and
  Nasrabadi}}]{Behkami02}
\bibinfo{author}{\bibfnamefont{A.~N.} \bibnamefont{Behkami}},
  \bibinfo{author}{\bibfnamefont{Z.}~\bibnamefont{Kargar}}, \bibnamefont{and}
  \bibinfo{author}{\bibfnamefont{N.}~\bibnamefont{Nasrabadi}},
  \bibinfo{journal}{Phys. Rev. C} \textbf{\bibinfo{volume}{66}},
  \bibinfo{pages}{064307} (\bibinfo{year}{2002}).

\end{thebibliography}

%\end{thebibliography}

\end{document}